\documentclass[twocolumn,preprintnumbers, 
               superscriptaddress,
               eqsecnum,nofootinbib,prd]{revtex4}
\usepackage{graphicx, fancybox}
\usepackage{amsmath,amssymb}
\usepackage[colorlinks=true, pdfstartview=FitV, linkcolor=red, citecolor=blue, urlcolor=blue]{hyperref}
\usepackage{color}
\usepackage{slashed}

\newcommand{\vect}[1]{\mathbf{#1}}

\newcommand{\re}{\mathrm{Re}}
\newcommand{\sgn}{\mathrm{sgn}}
\newcommand{\Slash}[1]{\ooalign{\hfil/\hfil\crcr$#1$}}

\newcommand{\vp}{\vect{p}}
\newcommand{\vk}{\vect{k}}
\newcommand{\vx}{\vect{x}}

\newcommand{\vv}{\vect{v}}

\newcommand{\vzero}{\vect{0}}
\newcommand{\vj}{\vect{j}}
\newcommand{\vA}{\vect{A}}
\newcommand{\vE}{\vect{E}}
\newcommand{\vB}{\vect{B}}

\newcommand{\Tr}{\mathrm{Tr}}

\newcommand{\comment}[1]{}

\newcommand{\Nc}{N_c}
\newcommand{\Nf}{N_f}
\newcommand{\Cf}{C_F}
\newcommand{\nf}{n_F}

\newcommand{\Cem}{C_{\text{em}}}

\begin{document}

\title{Finite temperature sum rules in the vector channel at finite momentum %\\
%and their application to lattice QCD data analysis 
}

\author{Philipp Gubler}
\email{pgubler@riken.jp}
\affiliation{Advanced Science Research Center, Japan Atomic Energy Agency, Tokai, Ibaraki 319-1195, Japan}
\affiliation{Department of Physics, Keio University, Kanagawa 223-8522, Japan}
\affiliation{Research and Education Center for Natural Science, Keio University, Kanagawa 223-8521, Japan}

\author{Daisuke Satow}
\email{dsato@th.physik.uni-frankfurt.de}
\affiliation{Institut f\"ur Theoretische Physik, Johann Wolfgang Goethe-Universit\"at,
Max-von-Laue-Str. 1, D-60438 Frankfurt am Main, Germany}

\begin{abstract} 
Exact sum rules for the longitudinal and transverse part of the vector channel spectral functions at nonzero momentum 
are derived in the first part of the paper. 
The sum rules are formulated for the finite temperature spectral functions, from which the vacuum component has been subtracted, 
and represent a generalization of previous work in which sum rules were derived only for the zero-momentum limit. 
In the second part of the paper, we demonstrate how the sum rules can be used as constraints in spectral fits to lattice data at various temperatures, 
with the latest dynamical lattice QCD data at zero momentum. 
\end{abstract} 

\date{\today}

\maketitle

%%%%%%%%%%%%%%%%%%%%%%%%%%%%%%%%%%%%%%%%%%%
\section{Introduction}
\label{sec:intro}

Hadrons, whose dynamics is governed by quantum chromodynamics (QCD), 
are deconfined at high temperature ($T$), 
where quarks and gluons are expected to be the fundamental degrees of freedom.
Such matter is called quark-gluon plasma, and is experimentally investigated by heavy ion collision experiments.
In analyzing the experimental data, electromagnetic (EM) probes such as dilepton spectra are particularly useful~\cite{Adare:2014fwh} 
because once generated in the medium, they are expected to reach the detector without further QCD interaction with other particles.
The electric conductivity is also an important quantity since it may increase the lifetime of magnetic fields generated in the early stage 
of the heavy ion collision~\cite{McLerran:2013hla, Tuchin:2013apa, Tuchin:2015oka, Hattori:2016emy}. 
Furthermore, how the spectrum of the vector meson is modified at finite $T$ has been discussed for a long time from the point of view of the 
chiral symmetry restoration~\cite{Bernard:1988db}.
The spectral function of the EM current at finite $T$, which is the focus of this paper, 
contains information on all the above three quantities. 

There are many approaches for evaluating the EM spectral function at finite $T$, 
such as perturbative QCD~\cite{Baier:1988xv}, holographic QCD~\cite{CaronHuot:2006te}, 
model calculations~\cite{Gale:2014dfa}, sum rules~\cite{Gubler:2015yna, Huang:1994fs, Kapusta:1993hq, Zschocke:2002mn,Kim:2017nyg}, 
low-energy effective theory~\cite{Chanfray,Klingl}, and lattice QCD~\cite{Ding:2010ga,Meyer:2011gj,Brandt:2015aqk, Amato:2013naa, Aarts:2007wj}. 
Nevertheless, none of these approaches are perfect.
Especially in the lattice QCD approach, which allows fully nonperturbative first principle QCD calculations, 
the spectral function cannot be analyzed directly since it is a quantity defined in real, not imaginary time.
Therefore, one needs to make some assumption about the functional form of the spectral function to analyze it, 
or otherwise has to rely on some method to analytically continue imaginary time data to real time, such as 
the maximum entropy method (MEM) \cite{Jarrell:1996rrw,Asakawa:2000tr,Gubler:2010cf}, 
the Backus-Gilbert method \cite{Backus:1968,Backus:1970,Brandt:2015sxa} or 
the Schlessinger point method \cite{Schlessinger:1967,Tripolt:2016cya} (see Ref.\,\cite{Tripolt:2017} for a comparison of these three 
methods). 

In our previous work of Ref.\,\cite{Gubler:2016hnf}, we derived three sum rules at zero momentum, which constrain the spectral function, 
and used them to improve the ansatz employed in previous lattice QCD analysis.
One aim of this paper is to derive similar sum rules for the small but finite spatial momentum case.
At finite momentum, the EM spectral function no longer has a %one
 single independent component, but two, corresponding to the transverse and the longitudinal channels.
In the longitudinal channel, a novel and robust structure, a sharp peak corresponding to the diffusion mode of the EM charge, appears. 

As shown in Ref.\,\cite{Gubler:2016hnf}, the sum rules can be used to improve the analysis of lattice QCD data by constraining the shape of the spectral functions and 
derive transport coefficients that do not appear directly in the spectral functions.
In Ref.\,\cite{Gubler:2016hnf}, this was demonstrated by using two sum rules (1 and 3 in this and the previous work), but the other sum rule (2) was not used. 
The other aim of this paper is to update the analysis such that all three sum rules can be used, 
and to employ the latest lattice QCD data including dynamical quarks as input for the spectral function fit. 

%%%%%%%%%
The paper is organized as follows:
In the next section, we introduce the quantities in quantum field theory that are necessary in our analysis, and explain how to derive the sum rules from 
the operator product expansion (OPE) and hydrodynamics~\cite{Romatschke:2009ng}.
Section~\ref{sec:sumrule} is devoted to the derivation of the sum rules in transverse and longitudinal channels, at small but finite spatial momentum. 
We also confirm that the spectral function evaluated at weak coupling and in the chiral limit satisfies these sum rules, and check to which energy region the sum rules are sensitive.
We demonstrate that the sum rules can be used to improve the lattice QCD analysis for the zero-momentum case in Sec.~\ref{sec:lattice}.
We summarize the paper and give concluding remarks in Sec.~\ref{sec:summary}.
In the three appendixes, we evaluate the contributions to the spectral functions from the transport peak, the continuum, and the UV tail, at weak coupling and in the chiral limit. 

In this paper, we recapitulate known results such as the sum rules in the transverse channel at zero momentum, Eqs.\,(\ref{eq:sumrule-1-p0}) and (\ref{eq:sumrule-3-T}). 
Their derivation can be found in our previous work~\cite{Gubler:2016hnf}, but we rederive them to make our paper self-contained. 
The recapitulation of the evaluation of the transport peak, the continuum, and the UV tails in the three appendixes is provided for the same reason.

%%%%%%%%%%%%%%%%%%%%%%%%%%%%%%%%%%%%%%%%%%%
\section{Preliminaries}

%What we do in this section
In this section, we explain the method for deriving the sum rules developed in Ref.~\cite{Romatschke:2009ng}.
Only the asymptotic behaviors of the EM current correlator in the UV and IR energy regions are necessary for this purpose. 
We also discuss these behaviors and give their analytic form 
obtained from OPE and hydrodynamics in this section. 

%%%%%%%%%%%%%%%%%%%%%%%%%%%%%%%%%%%%%%%%%%%
\subsection{Formalism}

We begin by introducing quantities that will be used in the derivation of the sum rules.
The retarded Green function of the EM current ($j^\mu\equiv e\sum_f q_f \overline{\psi}_f \gamma^\mu \psi_f $) is defined as 
$G^R_{\mu\nu}(\omega,\vp)\equiv i \int dt \int d^3\vx e^{i\omega t-i\vp\cdot \vx}
\theta(t) \langle [j_\mu(t,\vx),j_\nu(0,\vzero)] \rangle$, where the average is taken over the 
thermal ensemble, $e$ is the coupling constant of quantum electrodynamics, $q_f$ is the charge of each quark flavor (in units of $e$), 
and $\psi_f$ the quark field with flavor $f$, respectively.

At finite temperature, the medium effect breaks Lorentz symmetry so that the tensor structure of the Green function has two independent components, 
\begin{align}
\label{eq:tensor-decomposition}
G^R_{\mu\nu}(p)&=G^T(p)P^T_{\mu\nu}(p)
+G^L(p)P^L_{\mu\nu}(p),
\end{align}
where $p^\mu\equiv(\omega,\vp)$ is a shorthand notation for the energy and the spatial momentum, 
and $P^T_{\mu\nu}(p)\equiv g^{i}_\mu g^{j}_\nu (\delta^{ij}-{p^i p^j}/{\vp^2})$
and $P^L_{\mu\nu}(p)\equiv P^0_{\mu\nu}(p) -P^T_{\mu\nu}(p)$ are the projection operators to the transverse and longitudinal parts 
with $P^0_{\mu\nu}(p)\equiv -(g_{\mu\nu}-{p_\mu p_\nu}/{p^2})$.
The first (second) term in Eq.~(\ref{eq:tensor-decomposition}) corresponds to the transverse (longitudinal) component in three dimensions. 
When $\vp$ is along the $z$-direction, they are related to the components of the Green function as $G_{T}=G^R_{11}=G^R_{22}$ and $G_L=p^2G^R_{00}/\vp^2$.

\begin{figure}[t] 
\begin{center}
\includegraphics[width=0.25\textwidth]{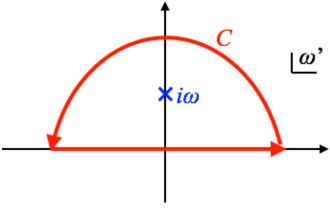} 
\vspace*{-1.5cm}
\caption{The contour $C$, used in the integral of Eq.~(\ref{eq:residue-theorem}).} 
\label{fig:contour} 
\end{center} 
\end{figure}

Here we recapitulate the method of deriving sum rules at finite temperature that was developed in Ref.\,\cite{Romatschke:2009ng}.
Using the residual theorem for the contour\footnote{This contour actually runs slightly above the real axis, so that it does not 
overlap with the singularities such as the continuum or the diffusion pole at zero momentum, which appear on it.}
 drawn in Fig.~\ref{fig:contour}, we get
\begin{align}
\label{eq:residue-theorem}
\begin{split}
&\delta G^{R}_{\mu\nu}(i\omega,\vp)-\delta G^{R}_{\mu\nu}(\infty,\vp) \\
&~~~= \frac{1}{2\pi i}\oint_C d\omega' 
\frac{\delta G^{R}_{\mu\nu}(\omega',\vp)-\delta G^{R}_{\mu\nu}(\infty,\vp)}{\omega'-i\omega},
\end{split}
\end{align}
where $\delta$ stands for the subtraction of the $T=0$ part: $\delta G^R_{\mu\nu}\equiv G^R_{\mu\nu}-G^R_{\mu\nu}|_{T=0}$.
Because of this subtraction of the zero temperature part and another subtraction of $\delta G^{R}_{\mu\nu}(\infty,\vp)$ done in the expression above, 
all the UV divergences are regularized in all the cases we consider.
Thus, the integral on the contour $C$ can be safely replaced with the integral on the real axis.
Moreover, in deriving Eq.~(\ref{eq:residue-theorem}) 
we have used the property that the retarded Green function is analytic in the upper $\omega'$ plane. 

Now, by taking the $\omega\rightarrow 0$ limit, Eq.~(\ref{eq:residue-theorem}) reduces to 
\begin{align}
\begin{split}
&\delta G^{R}_{\mu\nu}(0,\vp)-\delta G^{R}_{\mu\nu}(\infty,\vp) \\
&~~~= {\text P}\frac{1}{\pi i}\int^\infty_{-\infty} d\omega' 
\frac{\delta G^{R}_{\mu\nu}(\omega',\vp)-\delta G^{R}_{\mu\nu}(\infty,\vp)}{\omega'},
\end{split}
\end{align}
where we have used $1/(\omega'-i\omega)\rightarrow {\text P}(1/\omega')+i\pi\delta(\omega')$.
We consider only the case of $\mu=\nu$ in this paper.
Then, the real (imaginary) part of $\delta G^{R}_{\mu\nu}(\omega,\vp)$ is even (odd) in terms of $\omega$.
This property enables us to simplify the equation above as 
\begin{align}
\label{eq:derivation}
\begin{split}
\delta G^{R}_{\mu\nu}(0,\vp)-\delta G^{R}_{\mu\nu}(\infty,\vp) 
= \frac{2}{\pi}\int^\infty_{0} d\omega
\frac{\delta \rho_{\mu\nu}(\omega,\vp)}{\omega},
\end{split}
\end{align}
where we have introduced the spectral function\footnote{We note that our convention for $\rho_{\mu\nu}$ differs from the 
popular $\rho_{\mu\nu}=2{\text{Im}}G^R_{\mu\nu}$ by a factor of 2.} 
of the EM current, $\rho_{\mu\nu}(p)\equiv{\text{Im}}G^{R}_{\mu\nu}(p)$, and changed the label of the integration variable as $\omega'\rightarrow \omega$ for simplicity.
We hence see that the asymptotic behaviors of the retarded Green function in the UV and IR regions determine the integral of the spectral function. 

Let us here briefly discuss the differences between the sum rules derived here and 
the so-called finite energy sum rules (FESR), which are widely used in the literature 
(see for instance Ref.\,\cite{Ayala:2013vra}). In contrast to the procedure of this paper, one in the FESR 
does not subtract the zero temperature contribution as we have done in Eq.\,(\ref{eq:residue-theorem}) and below. 
Instead, to avoid an ultraviolet divergence, one does not take the radius of the contour in 
Fig.\,\ref{fig:contour} to infinity, but sets it to some threshold value, which however should be large enough 
such that the OPE is still approximately valid. Doing this, one can derive sum rules that 
are not exact, but practically useful, as they can constrain the spectral function below 
the threshold value that usually contains the most interesting physical content.

%%%%%%%%%%%%%%%%%%%%%%%%%%%%%%%%%%%%%%%%%%
%%%%%%%%%%%%%%%%%%%%%%%%%%%%%%%
\subsection{UV behavior}

The behavior in the UV region can be described with the help of the OPE.
At leading order in the coupling constant, the Wilson coefficients of the operators with dimension 4 read~\cite{CaronHuot:2009ns, Shifman:1978by}
\begin{align}
\label{eq:OPE}
\begin{split}
& \delta G^{R}_{\mu\nu}(\omega,\vp) \\
&=e^2\sum q^2_f \frac{1}{p^2} \\
&~~~\times \Bigl[\left\{2m_f \delta\left\langle\overline{\psi}_f \psi_f \right\rangle
+\frac{1}{12}\delta\left\langle \frac{\alpha_s}{\pi}G^2\right\rangle \right\}P^0_{\mu\nu}(p) \\
&~~~-2\delta\left\langle T^{\alpha\beta}_f \right\rangle A_{\mu\nu \alpha\beta}(p)
\Bigr]
+{\cal O}(\omega^{-4}), 
\end{split}
\end{align}
where $A_{\mu\nu \alpha\beta}(p)\equiv g_{\mu\alpha}g_{\nu\beta}+g_{\mu\beta}g_{\nu\alpha}-2(g_{\mu\alpha}p_{\nu}p_{\beta}+
g_{\nu\alpha}p_{\mu}p_{\beta}-g_{\mu\nu}p_\alpha p_\beta)/p^2$, 
$G^{\mu\nu}_a\equiv \partial^\mu A^\nu_a-\partial^\nu A^\mu_a -gf_{abc} A^\mu_b A^\nu_c$ is the field strength, 
$G^2\equiv G^a_{\mu\nu}G^{a\mu\nu}$, $T^{\alpha\beta}_{f}\equiv i{\cal {ST}} \overline{\psi}_f\gamma^\alpha D^\beta \psi_f $ is 
the quark component to the traceless part of the energy-momentum tensor, $D^\mu\equiv \partial^\mu +igA^\mu_a t^a$ is the covariant derivative, 
$A^\mu_a$ is the gluon field, $t^a$ is the generator of the $SU(\Nc)$ group in the fundamental representation, $f_{abc}$ is 
the structure constant of the $SU(\Nc)$ group, $m_f$ is the current quark mass, $g$ is the QCD coupling constant, $\alpha_s\equiv g^2/(4\pi)$, 
and $\Nc$ is the number of the colors. 
${\cal {ST}}$ makes a tensor symmetric and traceless, 
${\cal {ST}}O^{\alpha\beta}\equiv (O^{\alpha\beta}+O^{\beta\alpha})/2-g^{\alpha\beta}O^\mu_\mu /4$. 

We decompose Eq.\,(\ref{eq:OPE}) into transverse and longitudinal components as 
\begin{align}
\nonumber
\delta G_{T}(\omega,\vp) 
&=e^2\sum q^2_f \frac{1}{p^2} \\
\nonumber
&~~~\times \Bigl[\left\{2m_f \delta\left\langle\overline{\psi}_f \psi_f \right\rangle
+\frac{1}{12}\delta\left\langle \frac{\alpha_s}{\pi}G^2\right\rangle \right\}\\
\label{eq:OPE-wo-RG-T}
&~~~+\frac{8}{3}\frac{\omega^2+\vp^2}{p^2}\delta\left\langle T^{00}_f \right\rangle 
\Bigr]
+{\cal O}(\omega^{-4}),\\
\nonumber
\delta G^{R}_{00}(\omega,\vp) 
&=e^2\sum q^2_f \frac{1}{p^2}\frac{\vp^2}{p^2} \\ 
\nonumber
&~~~\times \Bigl[\left\{2m_f \delta\left\langle\overline{\psi}_f \psi_f \right\rangle
+\frac{1}{12}\delta\left\langle \frac{\alpha_s}{\pi}G^2\right\rangle \right\} \\
\label{eq:OPE-wo-RG-L}
&~~~+\frac{8}{3} \delta\left\langle T^{00}_f \right\rangle  
\Bigr]
+{\cal O}(\omega^{-6}),
\end{align}
where we have used the isotropy of the system and the traceless property of $T^{\alpha\beta}_f$.

Because we consider the $\omega\rightarrow \infty$ limit, 
we need to take into account the rescaling/mixing effect of the operators. 
Due to their vanishing anomalous dimensions, the chiral and gluon condensate terms remain unchanged, 
but the quark component of the energy-momentum tensor changes. 
To describe its behavior, we rewrite this operator as 
\begin{align}
\label{eq:Tf-decompose}
T^{00}_{f}=T'{}^{00}_{f}+\frac{1}{4\Cf+\Nf}\left(T^{00}+\frac{2}{\Nf}\tilde{T}^{00}\right),
\end{align} 
where 
\begin{align}
T'{}^{00}_{f}\equiv T^{00}_{f}-\frac{1}{\Nf}\sum_{f'} T^{00}_{f'},\\
T^{00}\equiv \sum_{f'} T^{00}_{f'}+T^{00}_g,\\
\tilde{T}^{00}\equiv 2C_F  \sum_{f'} T^{00}_{f'}-\frac{\Nf}{2} T^{00}_g .
\end{align} 
Here, $T^{\mu\nu}_g\equiv -G^{\mu\alpha}_{a}G^\nu{}_{\alpha a} +g^{\mu\nu}G^2/4$ is the gluon component of the traceless part of the energy-momentum tensor, 
$N_f$ is the flavor number, and $C_F\equiv (\Nc^2-1)/(2\Nc)$. 
We note that $T^{\mu\nu}$ is the traceless part of the full energy-momentum tensor, not the energy-momentum tensor itself.
A standard renormalization group (RG) analysis yields the following scaling properties~\cite{Peskin:1995ev}: 
\begin{align}
\label{eq:scaling-T'-Ttilde}
\begin{split} 
T'{}^{00}_{f}(\kappa)&=  \left[\frac{\ln\left(\kappa^2_0/\Lambda^2_{\text{QCD}} \right)}{\ln\left(\kappa^2/\Lambda^2_{\text{QCD}} \right)}\right]^{a'} 
T'{}^{00}_{f}(\kappa_0),\\
 \tilde{T}^{00}(\kappa)
&= \left[\frac{\ln\left(\kappa^2_0/\Lambda^2_{\text{QCD}} \right)}{\ln\left(\kappa^2/\Lambda^2_{\text{QCD}} \right)}\right]^{\tilde{a}}
 \tilde{T}^{00}(\kappa_0),
\end{split} 
\end{align}
while $T^{00}$ is independent of $\kappa$. 
Here $\kappa$ and $\kappa_0$ are renormalization scales, $\Lambda_{\text{QCD}}$ is the QCD scale parameter, 
$a'\equiv 8\Cf/(3b_0)$, and $\tilde{a}\equiv 2(4\Cf+\Nf)/(3b_0)$, where $b_0\equiv (11\Nc-2\Nf)/3$, 
which appears in the expression 
\begin{align}
\label{eq:RG-alphas}
\alpha_s(\kappa)=\frac{4\pi}{b_0 \ln(\kappa^2/\Lambda^2_{\text{QCD}})}.
\end{align}
In the $\omega\rightarrow \infty$ limit, it is natural to choose the RG 
scale\footnote{We could also choose $\kappa^2=p^2$, which however would not change the results of this paper.} 
as $\kappa^2=\omega^2$.

We see that, except for the $T^{00}$ term, all terms in Eq.~(\ref{eq:Tf-decompose}) are suppressed logarithmically at large $\omega$. 
Thus, Eqs.~(\ref{eq:OPE-wo-RG-T}) and (\ref{eq:OPE-wo-RG-L}) become
\begin{align} 
\nonumber
\delta G_{T}(\omega,\vp) 
&=e^2\sum q^2_f \frac{1}{p^2} \\
\nonumber
&~~~\times \Bigl[\left\{2m_f \delta\left\langle\overline{\psi}_f \psi_f \right\rangle
+\frac{1}{12}\delta\left\langle \frac{\alpha_s}{\pi}G^2\right\rangle \right\}\\
\label{eq:OPE-w-RG-T}
&~~~+\frac{8}{3}\frac{1}{4\Cf+\Nf}\frac{\omega^2+\vp^2}{p^2}\delta\left\langle T^{00} \right\rangle 
\Bigr] \nonumber \\
&~~~+{\cal O}(\omega^{-4}),\\
\nonumber
\delta G^{R}_{00}(\omega,\vp) 
&=e^2\sum q^2_f \frac{1}{p^2}\frac{\vp^2}{p^2} \\ 
\nonumber
&~~~\times \Bigl[\left\{2m_f \delta\left\langle\overline{\psi}_f \psi_f \right\rangle
+\frac{1}{12}\delta\left\langle \frac{\alpha_s}{\pi}G^2\right\rangle \right\} \\
\label{eq:OPE-w-RG-L}
&~~~+\frac{8}{3}\frac{1}{4\Cf+\Nf} \delta\left\langle T^{00} \right\rangle  
\Bigr]
+{\cal O}(\omega^{-6}). 
\end{align} 
We note that, in the $\omega\rightarrow \infty$ limit, which is relevant to the derivation of the sum rule, 
the asymptotic freedom of QCD guarantees that the above expression is exact. 
In other words, all higher order $\alpha_s$ corrections vanish in this limit. 

%%%%%%%%%%%%%%%%%%%%%%%%%%%%%%%
\subsection{IR behavior}
\label{ssc:IR}

On the other hand, the asymptotic behavior in the IR region is described by hydrodynamics~\cite{Kovtun:2012rj}, as long as the spatial momentum is small enough.
In the channel of the EM current, the basic equations consist of the conservation law and the constituent equation, 
\begin{align}
\label{eq:conservation-law}
\partial_0j^0&= -\nabla\cdot \vj,\\
\nonumber
\vj&= -D\nabla j^0+\sigma\vE
-\sigma\tau_J\partial_0 \vE+\kappa_B\nabla\times\vB  \\ 
\label{eq:constituent-eq}
&~~~+{\cal O}(\partial^2 \vE, \partial^2 \vB, \partial^2 j^0 ),
\end{align}
where $D$ is the diffusion constant, $\sigma$ the electrical conductivity and $\tau_J$ and $\kappa_B$ second order transport 
coefficients corresponding to $\partial_0\vE$ and $\nabla\times\vB$, respectively.
$\vE\equiv -\nabla A^0-\partial_0 \vA$ and $\vB\equiv\nabla\times\vA$ are the electric and magnetic fields, where $A^\mu$ is the vector potential. 

After performing the Fourier transformation that is defined as $f(p)\equiv \int d^4x e^{ip\cdot x} f(x)$, 
Eqs.~(\ref{eq:conservation-law}) and (\ref{eq:constituent-eq}) become
\begin{align} 
\label{eq:conservation-law-p}
\omega j^0&= \vp\cdot \vj,\\
\nonumber
\vj&= -Di\vp j^0+i\sigma (-\vp A^0+\omega \vA)\\
\nonumber
&~~~-\sigma\tau_J \omega  (-\vp A^0+\omega \vA)-\kappa_B ([\vp\cdot\vA]\vp-\vp^2\vA) \\
\label{eq:constituent-eq-p} 
&~~~+{\cal O}(p^2A, p^2j^0).
\end{align}
Let us solve these equations for the transverse and the longitudinal components of the current.
By introducing the transverse component, $j^i_T(p)\equiv P^{ij}_T(p) j^j(p)$, we get the solutions as follows:
\begin{align}
\vj_T(p)&= \left(i\sigma\omega-\sigma\tau_J\omega^2+\kappa_B\vp^2
\right) \vA_T \nonumber \\
&~~~ +{\cal O}(\omega^3\vA_T, \omega\vp^2\vA_T, \vp^4\vA_T),\\
\nonumber
\omega j^0(p)&= -iD\vp^2 j^0
-i\vp^2\sigma A^0 \\
&~~~+ {\cal O}(\omega\vp^2 A^0, \vp^4 A^0, \omega\vp^2 j^0, \vp^4 j^0) \nonumber \\
&~~~+({\text {terms that are proportional to }}\vp\cdot\vA).
\end{align} 

By using the linear response theory, the induced current is written as 
\begin{align}
\label{eq:linear-response}
j_\mu(p)=-G^R_{\mu\nu}(p)A^\nu(p),
\end{align}
 from which we obtain
\begin{align}
\label{eq:hydro-GR-T}
G_{T}(p) &= i\sigma\omega-\sigma\tau_J\omega^2+\kappa_B\vp^2
+{\cal O}(\omega^3, \omega\vp^2, \vp^4),\\
\label{eq:hydro-GR-L}
G^R_{00}(p) &= i\sigma\vp^2\frac{1+{\cal O}(\omega,\vp^2)}{\omega+iD\vp^2+{\cal O}(\omega\vp^2,\vp^4)}.
\end{align}
We note that there is a pole at $\omega=-iD\vp^2$ in the longitudinal channel, which we call the diffusion pole.
This is a novel structure that appears only at finite $\vp$.
This pole appears as a peak in the spectral function,
\begin{align}
\label{eq:hydro-rho-L}
\rho_{00}(p) &= \sigma\vp^2
\frac{\omega}{\omega^2+(D\vp^2)^2} ,
\end{align}
while at $|\vp|=0$, it reduces to the delta function, 
\begin{align}
\label{eq:hydro-rho-L-p=0}
\rho_{00}(\omega,\vzero) 
&= \pi\frac{\sigma}{D}\omega\delta(\omega) .
\end{align}

For constructing our sum rules, we, in principle, also 
need to evaluate the zero temperature part, which should be subtracted later. 
At $T=0$, Lorentz invariance guarantees the tensor structure of the correlator, 
\begin{align}
G^R_{\mu\nu}(p)
&= p^2P^0_{\mu\nu}(p) \tilde{G}^R(p^2),
\end{align}
where $\tilde{G}^R(0)=0$ due to the renormalization condition of the electric charge~\cite{Peskin:1995ev}. 
The transverse component is given as $G_T(p)= p^2\tilde{G}^R(p^2)$.
%The expansion of $\tilde{G}^R(p^2)$ at small $p^2$ starts from $p^2$ term 
$\tilde{G}^R(p^2)$ is regular at $p^2=0$ due to the renormalization condition, so it is easy to see that there are no 
contributions to $\sigma$, $\tau_J$, and $\kappa_B$ at $T=0$, from Eq.~(\ref{eq:hydro-GR-T}). 
The $T=0$ contribution only enters in the higher order terms, which are neglected in Eq.~(\ref{eq:hydro-GR-T}). 

On the other hand, the longitudinal component reads $G^R_{00}(p)=\vp^2 \tilde{G}^R(p^2)$, so $G^R_{00}(p)$ at 
$T=0$ and $\omega=0$ is of order $\vp^2 \tilde{G}^R(-\vp^2)$.
The $T=0$ contribution does not affect the sum rules 2 (\ref{eq:sumrule-2-L}) and 3 (\ref{eq:sumrule-3-L}) because 
as is explained in Sec.~\ref{ssc:sumrule-L}, the relevant quantities to the derivation of these sum rules 
are $\omega^2\delta G^R_{00}(p)|_{\omega\rightarrow 0}$ and $\omega^4\delta G^R_{00}(p)|_{\omega\rightarrow 0}$, 
and the $T=0$ contributions vanish. 
The only sum rule which $T=0$ terms may affect is sum rule 1 (\ref{eq:sumrule-1-L}).
Nevertheless, we only consider terms in $G^R_{00}(p)|_{\omega\rightarrow 0}$ up to $\vp^2$ in this paper, as is shown in Sec.~\ref{ssc:sumrule-L}.
Therefore, sum rule 1 is also unaffected.

%at T=0, do we have spectrum below 2-pi channel?->NO.

We note that we have so far neglected effects of possible hydro modes, which adds a $\omega^{3/2}$ term at zero momentum to $G^T$~\cite{Kovtun:2003vj}. 
It was suggested that this approximation is justified in the large $\Nc$ limit~\cite{Kovtun:2003vj}.
Among our sum rules, sum rule 3 (\ref{eq:sumrule-3-T}) in the transverse channel may be changed at finite $\Nc$ due to this effect.
In the analysis of lattice QCD data, this effect practically can be neglected because the IR cutoff of the current lattice QCD is not small enough for such IR energy effects to be detectable. 

%%%%%%%%%%%%%%%%%%%%%%%%%%%%%%%%%%%%%%%%%%%
\section{Sum rules at finite momentum}
\label{sec:sumrule}

In this section, we derive the sum rules at finite momentum in the transverse and the longitudinal channels.
For later convenience, we also give the expressions for the sum rules at zero momentum in the transverse channel, which were already obtained in Ref.~\cite{Gubler:2016hnf}.
We also confirm that the sum rules are satisfied by the expressions of the spectral function in the chiral and weak coupling limits.

%%%%%%%%%%%%%%%%%%%%%%%%
\subsection{Transverse channel}

\begin{figure}[t] 
\begin{center}
%\vspace*{-1cm}
%\hspace*{1.2cm}
\includegraphics[width=0.49\textwidth]{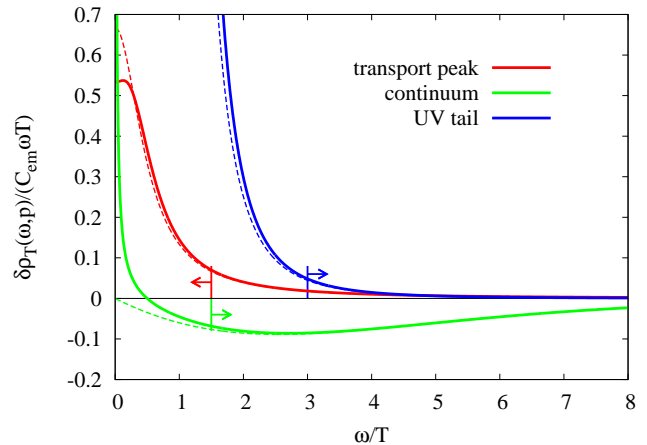} 
%\vspace*{0.5cm}
\caption{
The transport peak, the continuum, and the UV tail of the EM current spectral function in the transverse channel $\delta\rho_T$ as a function of $\omega$. 
The energy unit is $T$. 
To draw the figure, the parameters are set as $\Nc=\Nf=3$, $\tau^{-1}/T=0.5$, $\kappa_0/T=1$, and $\Lambda_{\text{QCD}}/T=0.67$. 
The solid (dashed) lines correspond to a spatial momentum of $|\vp|/T = 0.5$ ($|\vp| = 0$). The vertical lines with 
the attached arrows indicate the regions for which the respective analytic expressions can be trusted.} 
\label{fig:spectral-longpaper} 
\end{center} 
\end{figure}

Before deriving the sum rules, let us discuss what kind of structure can be expected to appear in the spectral function in a perturbative analysis.
First, in the low-energy region, a peak with a width of order $g^4T$ due to the collision effect is expected to appear.
This peak is called the transport peak.
Its derivation is recapitulated in Appendix~\ref{app:transport-peak}.
At larger energy, $\omega\sim T$, the pair-creation process yields a continuum in the spectral function (see Appendix~\ref{app:continuum} for its expression).
Also, at $\omega\gg T$, the OPE analysis predicts a UV tail, whose derivation is recapitulated in Appendix~\ref{app:UVtail}.
These structures are summarized in Fig.~\ref{fig:spectral-longpaper}, where the expressions in the chiral and weak 
coupling limits for $\Nc=\Nf=3$, Eqs.~(\ref{eq:transport-T}), (\ref{eq:continuum-T}), and (\ref{eq:UVtail-T}) have been used. 
The following parameters are chosen for illustrative purposes: 
$\tau^{-1}/T=0.5$, $|\vp|/T=0.5$, $\kappa_0/T=1$, and $\Lambda_{\text{QCD}}/T=0.67$. 
We note that the corrections due to the $\vp^2$ terms are almost negligible for this case, 
though the $|\vp|$ value adopted here is not very small compared to $T$. 
We furthermore caution that the plots for each structure are reliable only at their energy regions of applicability. 
Namely, the transport peak is reliable at low energy, the continuum at intermediate and high energy and the UV tail at high energy, respectively. 
These regions are marked by the vertical lines with attached arrows in the figure. Note that 
these boundaries are not exact and should only be considered as indicative. 
One should not take the curves seriously when they are outside of the adequate energy regions. 

%%%%%%%%%%%%%%%%%%%%%%

%%%%%%%%%%%%%%
\subsubsection{Sum rule 1}

By using the asymptotic expression of $\delta G^R_T$ in the UV and IR energy regions, Eqs.~(\ref{eq:OPE-w-RG-T}) and (\ref{eq:hydro-GR-T}), Eq.~(\ref{eq:derivation}) for $\mu=\nu=1$ becomes 
\begin{align}
\label{eq:sumrule-1-T}
\begin{split}
\kappa_B \vp^2
+{\cal O}(\vp^4)
= \frac{2}{\pi}\int^\infty_{0} d\omega
\frac{\delta \rho_{T}(\omega,\vp)}{\omega},
\end{split}
\end{align} 
where the contribution from the UV part has vanished.
This is the first sum rule in the transverse channel (sum rule 1).
We note that $|\vp|$ should be small enough to trust this relation, because we have assumed that the IR region 
is well described by hydrodynamics, which is valid only at small momentum and energy. 
$\kappa_B$ has been evaluated in lattice QCD~\cite{Brandt:2013faa} with methods that do not 
suffer from the problem 
of analytic continuation, so this sum rule can be used to constrain the spectral function.
At $|\vp|=0$, it reduces to
\begin{align}
\label{eq:sumrule-1-p0}
\begin{split}
0&= \frac{2}{\pi}\int^\infty_{0} d\omega
\frac{\delta \rho_{T}(\omega,\vzero)}{\omega},
\end{split}
\end{align} 
which was already obtained in Ref.~\cite{Gubler:2016hnf}, and also in Ref.~\cite{Bernecker:2011gh} by using the conservation of the EM current.

To get a feeling of how this sum rule is satisfied, let us 
check the respective contributions from the transport peak, the continuum, and the UV tail below.
This also gives us an indication about the sensitivity of the sum rule integral to these three structures.
The transport peak in the spectral function at small momentum is given by Eq.~(\ref{eq:transport-T}), and its contribution to sum rule 1 reads
\begin{align}
\label{eq:T-sumrule1-transport}
\begin{split}
 \frac{2}{\pi}\int^\infty_{0} d\omega
\frac{\delta \rho_{T}(\omega,\vp)}{\omega}
&\simeq  \Cem\Nc\chi \frac{2}{3} \frac{2}{\pi}
\int^\infty_{0} d\omega \frac{\tau^{-1} }{\omega^2+\tau^{-2}} \\
&~~~\times \left[1 +\frac{\vp^2}{5} \frac{(3\omega^2-\tau^{-2})}{(\omega^2+\tau^{-2})^2}\right]\\
&= \Cem\Nc\chi \frac{2}{3},
\end{split}
\end{align}
where $\Cem\equiv e^2\sum_f q^2_f$, $\tau\sim (g^4T)^{-1}$ is the relaxation time introduced in the Boltzmann equation, 
and $\chi\equiv T^2/6$. For the leading term we have used $\int^\infty_{0} d\omega \tau^{-1}/(\omega^2+\tau^{-2})=\pi/2$, 
while the momentum dependent term vanishes because of 
$\int^\infty_{0} d\omega (3\omega^2-\tau^{-2})/(\omega^2+\tau^{-2})^3=-[\omega/(\omega^2+\tau^{-2})^2]^\infty_0=0$.

The contribution from the continuum (\ref{eq:continuum-T}) to sum rule 1 is given as 
\begin{align}
\label{eq:T-sumrule1-cont}
\begin{split}
& \frac{2}{\pi}\int^\infty_{0} d\omega
\frac{\delta \rho_{T}(\omega,\vp)}{\omega} \\
&\simeq -C_{\text{em}}\Nc  \frac{1}{3\pi^2}
\int^\infty_{0} d\omega \omega
 \Biggl[\nf\left(\frac{\omega}{2}\right) \\
&~~~ +\vp^2\left\{\frac{1}{20}\nf''\left(\frac{\omega}{2}\right)
 -\frac{1}{\omega^2}\nf\left(\frac{\omega}{2}\right)\right\}\Biggr] \\
 &\simeq -C_{\text{em}}\Nc  \frac{1}{3}
 \Biggl[\frac{T^2}{3} 
 -\frac{\vp^2}{2\pi^2}\ln\frac{T}{\mu}\Biggr] ,
\end{split}
\end{align}
where  $\nf(k^0)\equiv[e^{k^0/T}+1]^{-1}$ is the Fermi distribution function. 
We have used $\int^\infty_{0} d\omega \omega \nf (\omega/2)=\pi^2T^2/3$, and introduced the IR cutoff $\mu$ for the continuum contribution because of a logarithmic IR divergence.
The nonsingular parts of the $\vp^2$ terms have been omitted.
The result for the continuum (\ref{eq:continuum-T}) is obtained from a one-loop calculation, 
and becomes unreliable when $\omega\lesssim gT$, where the hard-thermal loop resummation becomes necessary~\cite{Moore:2006qn, Braaten:1990wp}.
Therefore, we see that $\mu\sim gT$.

The contribution from the UV tail (\ref{eq:UVtail-T}) is estimated as
\begin{align}
\label{eq:T-sumrule1-UVtail}
\begin{split}
 \frac{2}{\pi}\int^\infty_{0} d\omega
\frac{\delta \rho_{T}(\omega,\vp)}{\omega}
&\sim \Cem g^2T^2, \Cem g^2\vp^2.
\end{split} 
\end{align}
Here we have used the fact that the IR cutoff of the UV tail is of order $\sim T$, because the derivation of the UV tail is based on the OPE, which is valid when $\omega\gg T$.

Now let us check that sum rule 1 (\ref{eq:sumrule-1-T}) is satisfied.
Because $\kappa_B=0$ at weak coupling (see Appendix~\ref{app:transport-peak}), the integral in Eq.~(\ref{eq:sumrule-1-T}) needs 
to vanish in order to satisfy the sum rule.
We first see that for the $\vp$-independent part, the contributions from the transport peak and the continuum, 
which are of order $\Cem T^2$, cancel while the contribution from the UV tail is of higher order ($\sim \Cem g^2T^2$), 
by looking at Eqs.\,(\ref{eq:T-sumrule1-transport}-\ref{eq:T-sumrule1-UVtail}). 
Therefore, the $\vp$-independent part was shown to satisfy sum rule 1 at leading order already in Ref.\,\cite{Gubler:2016hnf}. 
For the $\vp^2$ term, we see that the transport peak does not contribute, and the continuum contribution is of order $\sim \Cem \vp^2\ln (1/g)$ 
while the UV tail contribution is suppressed by a factor of $g^2$.
Because the continuum contribution is sensitive to the IR cutoff, we need to improve the evaluation by performing the HTL resummation~\cite{Moore:2006qn, Braaten:1990wp}, 
in order to confirm that this contribution becomes negligible so that sum rule 1 is satisfied at order $\vp^2$. 
It is furthermore understood that sum rule 1 is mainly sensitive to the transport peak as well as the continuum, while the contribution of the UV tail is small.

%%%%%%%%%%%%%%
\subsubsection{Sum rule 2}
\label{sssec:sumrule2-T}

In the derivation of Eq.\,(\ref{eq:derivation}), we used only the fact that the retarded Green function is analytic in the upper $\omega$ plane.
Thus, we can derive a similar equation in which $\delta G^R(\omega)$ ($\delta \rho(\omega)$) is replaced 
with $\omega^2\delta G^R(\omega)$ ($\omega^2\delta \rho(\omega)$), 
\begin{align}
\begin{split}
&\omega^2\delta G_{T}(\omega,\vp)|_{\omega\rightarrow 0}-\omega^2\delta G_{T}(\omega,\vp)|_{\omega\rightarrow \infty} \\
&= \frac{2}{\pi}\int^\infty_{0} d\omega \omega
\delta \rho_{T}(\omega,\vp),
\end{split}
\end{align}
for the transverse component.
By using Eqs.~(\ref{eq:OPE-w-RG-T}) and (\ref{eq:hydro-GR-T}), this equation becomes
\begin{align}
\label{eq:sumrule-2-T}
\begin{split}
&-e^2\sum q^2_f 
\Biggl[\left\{2m_f \delta\left\langle\overline{\psi}_f \psi_f \right\rangle
+\frac{1}{12}\delta\left\langle \frac{\alpha_s}{\pi}G^2\right\rangle \right\}\\
&~~~+\frac{8}{3}\frac{1}{4\Cf+\Nf}\delta\left\langle T^{00} \right\rangle 
\Biggr] 
= \frac{2}{\pi}\int^\infty_{0} d\omega \omega
\delta \rho_{T}(\omega,\vp). 
\end{split}
\end{align}
We call this equation sum rule 2 in what follows. 
It should be noted that there is no explicit $|\vp|$ dependence on the left-hand side, but again we are implicitly 
assuming that $|\vp|$ is small enough so that the hydrodynamics well describes the behavior of $\omega^2\delta G_{T}(\omega,\vp)|_{\omega\rightarrow 0}$.
Also, we emphasize that the expectation values of the local operators on the left-hand side can be evaluated 
nonperturbatively by 
lattice QCD without suffering from the problem of analytic continuation. 
Therefore, sum rule 2 can be used to constrain the shape of the spectral function.
As it was already discussed in the previous paper~\cite{Gubler:2016hnf}, lattice QCD results show that 
the left-hand side of Eq.\,(\ref{eq:sumrule-2-T}) is found to be  dominated 
by the $\left\langle T^{00} \right\rangle$ for almost all temperatures around and above $T_c$. 

Let us evaluate the contributions to sum rule 2 from the transport peak, continuum, and UV tail.
The contribution from the transport peak is found by using Eq.~(\ref{eq:transport-T}) as 
\begin{align}
\label{eq:T-sumrule2-transport}
\begin{split}
\frac{2}{\pi}\int^\infty_{0} d\omega \omega
\delta \rho_{T}(\omega,\vp)
&=  \Cem\Nc\chi \frac{2}{3} \frac{2}{\pi}\int^\infty_{0} d\omega 
  \frac{\tau^{-1} \omega^2}{\omega^2+\tau^{-2}} \\
&~~~\times \left[1 +\frac{\vp^2}{5} \frac{(3\omega^2-\tau^{-2})}{(\omega^2+\tau^{-2})^2}\right]\\
 &= \Cem\Nc\chi \frac{2}{3}
 \left[ \frac{2}{\pi}\tau^{-1}\varLambda +\frac{\vp^2}{5} \right] ,
 \end{split}
\end{align}
where we have introduced the UV cutoff $\varLambda$ for the transport peak because of the linear UV divergence, 
and used $\int^\infty_{0} d\omega \omega^2(3\omega^2-\tau^{-2})/(\omega^2+\tau^{-2})^3=\pi \tau/2$. 
As the Boltzmann equation cannot be used when $\omega\gtrsim gT$ since the instantaneous scattering description 
becomes invalid~\cite{Moore:2006qn}, we set $\varLambda\sim gT$.

The contribution from the continuum (\ref{eq:continuum-T}) reads
\begin{align}
\label{eq:T-sumrule2-cont}
\begin{split}
& \frac{2}{\pi}\int^\infty_{0} d\omega \omega
\delta \rho_{T}(\omega,\vp) \\
&=  -  C_{\text{em}}\Nc \frac{1}{3\pi^2} 
\int^\infty_{0} d\omega \omega^3
 \Biggl[\nf\left(\frac{\omega}{2}\right) \\
&~~~~ +\vp^2\left\{\frac{1}{20}\nf''\left(\frac{\omega}{2}\right)
 -\frac{1}{\omega^2}\nf\left(\frac{\omega}{2}\right)\right\}\Biggr] \\
 &= -  C_{\text{em}}\Nc \frac{1}{45} T^2
 \Biggl[ 14\pi^2T^2  +\vp^2 \Biggr] ,
\end{split}
\end{align}
where we have used $\int^\infty_{0} d\omega \omega^3 \nf (\omega/2)=14\pi^4T^4/15$ and $\int^\infty_{0} d\omega \omega^3 \nf'' (\omega/2)=8\pi^2T^2$. 

The UV tail contribution is, by using Eq.~(\ref{eq:UVtail-T}), found to be
\begin{align}
\label{eq:T-sumrule2-UVtail}
\begin{split}
 \frac{2}{\pi}\int^\infty_{0} d\omega \omega
\delta \rho_{T}(\omega,\vp)
&=  \frac{2}{\pi} \Cem \Nc\Cf \frac{4\pi^2T^4}{27}   \alpha_s(\kappa_0)  \\
&~~~\times \int^\infty_{X_0} dX
 \left[\frac{X_0}{X}\right]^{\tilde{a}+1} \\
&~~~+{\cal O}\left(\Cem g^2T^2\vp^2\right)\\
&=  \Cem \Nc\Cf \frac{4\pi^2T^4}{9}   \frac{2}{4\Cf+\Nf}   \\
&~~~+{\cal O}\left(\Cem g^2T^2\vp^2\right),
\end{split}
\end{align}
where we have introduced $X\equiv \ln(\omega/\Lambda_{\text{QCD}})$ and $X_0\equiv \ln(\kappa_0/\Lambda_{\text{QCD}})$, 
used Eq.~(\ref{eq:RG-alphas}) and introduced the IR cutoff of the UV tail as $\kappa_0\sim T$.

Let us check whether sum rule 2 is satisfied.
From Eqs.\,(\ref{eq:T-sumrule2-transport}-\ref{eq:T-sumrule2-UVtail}), we see that for the $\vp$-independent part, 
the contributions from the continuum and the UV tail have the same order of magnitude ($\sim \Cem T^4$) while the contribution from the transport peak is 
much smaller, $\sim \Cem T^2 \tau^{-1}\varLambda\sim \Cem g^5 T^4$. 
These contributions are found to agree with the left-hand side of Eq.~(\ref{eq:sumrule-2-T}) by using $\delta \langle T^{00}\rangle=\Nc\pi^2T^4(8\Cf+7\Nf)/60$, 
which is obtained from Eqs.~(\ref{eq:T00f-free}) and (\ref{eq:T00g-free}). 
For the $\vp^2$ term, the contributions from the transport peak and the continuum cancel, while the UV tail contribution is much smaller. 
Therefore, we have confirmed that sum rule 2 is satisfied up to order $\vp^2$, in the chiral and weak coupling limits.

%%%%%%%%%%%%%%
\subsubsection{Sum rule 3}

The derivation of the third sum rule turns out to be somewhat more tricky.
Equation~(\ref{eq:residue-theorem}) for $\mu=\nu=1$ can be rewritten as
\begin{align}
\begin{split}
&\delta G_{T}(i\omega,\vp)-\delta G_{T}(\infty,\vp) 
= \frac{1}{\pi }\int^\infty_{0} d\omega' \\
&~~~\times\frac{\omega' \delta \rho_{T}(\omega',\vp)+\omega[\re\delta G_{T}(\omega',\vp)-\delta G_{T}(\infty,\vp)]}{\omega'{}^2+\omega^2}\\
&~~~= \frac{2}{\pi }\int^\infty_{0} d\omega' 
\frac{\omega' \delta \rho_{T}(\omega',\vp)}{\omega'{}^2+\omega^2} ,
\end{split}
\end{align}
where we have used the relation 
\begin{align}
&0 = \nonumber \\
& \int^\infty_{-\infty} d\omega' 
\frac{\omega' \delta \rho_{T}(\omega',\vp)-\omega[\re\delta G_{T}(\omega',\vp)-\delta G_{T}(\infty,\vp)]}{\omega'{}^2+\omega^2},
\end{align}
 in the last line, which is obtained by using the residual theorem for the integral $\oint_C d\omega' 
[\delta G_{T}(\omega',\vp)-\delta G_{T}(\infty,\vp)]/(\omega'+i\omega)$.
By subtracting Eq.~(\ref{eq:derivation}) and $i\omega \delta G_T{}'=i\omega^2\delta G_T{}'2\int^\infty_0 d\omega'/[\pi(\omega^2+\omega'{}^2)]$, 
which is necessary to regularize the IR singularity in the integral, we get 
\begin{align}
\begin{split}
&\delta G_{T}(i\omega,\vp)-\delta G_{T}(0,\vp) -i\omega \delta G_T{}'(0,\vp) \\
&= \frac{2}{\pi }\omega^2 \int^\infty_{0} d\omega' \frac{1}{\omega^2+\omega'{}^2}
\left[ \delta \rho_{T}(\omega',\vp)\frac{-1}{\omega'} 
+\delta \rho'_T{}(0,\vp)\right],
\end{split}
\end{align}
where $'$ stands for the derivative in terms of energy ($\omega$, $\omega'$). 
Taking the $\omega \to 0$ limit, this reduces to
\begin{align}
\label{eq:derivation-sumrule3-T}
\begin{split}
\frac{1}{2}\delta G_{T}{}''(0,\vp)
= \frac{2}{\pi } \int^\infty_{0} d\omega \frac{1}{\omega^3}
\left[ \delta \rho_{T}(\omega,\vp)
-\omega \delta \rho'_T{}(0,\vp)\right]. 
\end{split}
\end{align}
Here, we have changed the integration variable from $\omega'$ to $\omega$ for simplicity.
To get an explicit form of this sum rule, one needs to evaluate $\delta G_{T}{}''(0,\vp)$ and $\delta \rho'_T{}(0,\vp)$. 
%Naively, it seems that these quantities can be obtained by expanding Eq.~(\ref{eq:hydro-GR-T}) in terms of $|\vp|$ when $\vp$ is small enough:
%For example, 
In the expansion of Eq.\,(\ref{eq:hydro-GR-T}), we get only the $|\vp|=0$ terms as $\delta G_{T}{}''(0,\vp)=-2\sigma\tau_J+{\cal O}(\vp^2)$ and $\delta \rho'_T{}(0,\vp)=\sigma+{\cal O}(\vp^2)$.
%Unfortunately, we can not truncate the series even when $|\vp|$ is small, because however small the truncated term to the integrand of the integral in Eq.~(\ref{eq:derivation-sumrule3-T}) is, it causes the IR divergence in the integral.
Therefore, we can obtain a sum rule for the $|\vp|=0$ case, which reads
\begin{align}
\label{eq:sumrule-3-T}
-\sigma\tau_J
&= \frac{2}{\pi} \int^\infty_{0} \frac{d\omega}{\omega^3} 
\left[\delta \rho_T(\omega,\vzero)-\sigma\omega\right].
\end{align}
We call this equation sum rule 3 for the transverse channel.
We note that the transport coefficients in the left-hand side cannot be computed by lattice QCD without suffering from the problem of analytic continuation.
We do not check that sum rule 3 is satisfied in the chiral and the weak coupling limits since this was already done in our previous paper~\cite{Gubler:2016hnf}.
Instead we just cite the order of magnitude of the three contributions:
the transport peak contribution is of order $\Cem T^2 \tau^2\sim \Cem g^{-8}$ and is equal to the left-hand side of sum rule 3, while the continuum is much smaller, $\Cem g^{-5}$.
The UV tail contribution is the smallest and of order $\Cem g^{-4}$. 

Explicitly taking into account higher order terms in Eq.~(\ref{eq:hydro-GR-T}) and expanding $\delta\rho_T$ in $|\vp|^2$ order by order, 
it should be possible to obtain a corresponding sum rule at finite momentum. 
We leave this task for future work. 

%%%%%%%%%%%%%%%%%%%%%%%%
\subsection{Longitudinal channel}
\label{ssc:sumrule-L}

Before discussing the sum rules, let us remember that the retarded Green function in the longitudinal channel is exactly known 
at zero momentum~\cite{Ding:2010ga} from the conservation law of the charge:
\begin{align}
\label{eq:rho-L-p=0-exact}
\rho_{00}(\omega,\vzero) 
&= \pi \chi_q \omega\delta(\omega) ,
\end{align}
where $\chi_q\equiv \int d^3\vx \langle j^0(\vx) j^0(\vzero) \rangle/T%
$ is the charge susceptibility.
By matching this result with the hydro result of Eq.\,(\ref{eq:hydro-rho-L-p=0}), we see that the hydro result is exact for all $\omega$ at zero momentum, and $\sigma/D=\chi_q$.
For this reason, the sum rules in the longitudinal channel provide nontrivial information only when $\vp$ is finite.
Therefore, we consider only the finite momentum case in this subsection. 

\begin{figure}[t] 
\begin{center}
%\vspace*{-1cm}
%\hspace*{1.2cm}
\includegraphics[width=0.49\textwidth]{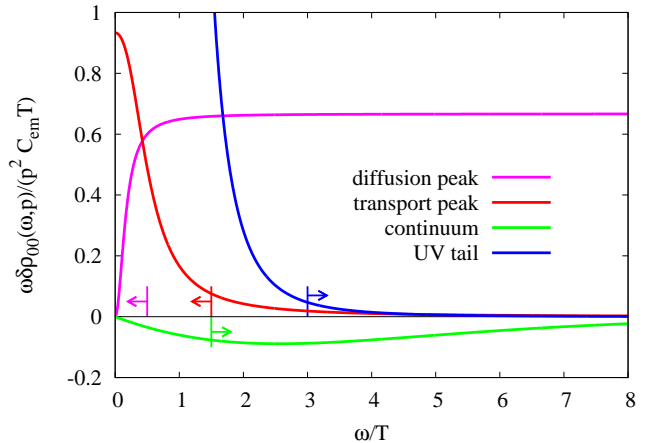} 
%\vspace*{0.5cm}
\caption{
The transport peak, the continuum, and the UV tail of the EM current spectral function in the longitudinal channel $\delta\rho_{00}$ as a function of $\omega$. 
The energy unit is $T$. 
The parameters and the meaning of the vertical line with attached arrows are the same as in Fig.~\ref{fig:spectral-longpaper}.} 
\label{fig:spectral-longpaper-L} 
\end{center} 
\end{figure}

At finite momentum, the diffusion peak appears in the longitudinal spectral function in addition to the three structures 
that were already present in the transverse channel, as was explained in Sec.~\ref{ssc:IR}. 
To get a feeling about the possible shape of the spectral function in the longitudinal channel, we plot the diffusion peak (\ref{eq:hydro-rho-L}), 
the transport peak (\ref{eq:transport-00}), the continuum (\ref{eq:continuum-00}), and the UV tail (\ref{eq:UVtail-00}) in Fig.~\ref{fig:spectral-longpaper-L}. 
The parameters are the same as in Fig.~\ref{fig:spectral-longpaper}. 
Again, the approximate regions for which the above analytic descriptions are expected to be valid are indicated by the vertical lines and arrows. 

Here we comment on the treatment of the diffusion peak in the traditional QCD sum rule literature: 
In the conventional sum rule approach, the delta function structure that is similar to the hydro result of Eq.\,(\ref{eq:hydro-rho-L-p=0}) was 
suggested based on the perturbative calculation in Ref.\,\cite{Bochkarev:1985ex}, and has been assumed in the subsequent works. 
Though the two approaches give the same form at $|\vp|=0$ as they should follow the exact results (\ref{eq:rho-L-p=0-exact}), 
the perturbative approach is not generally reliable at $\omega=|\vp|=0$ even when $g$ is small, so that the hydro approach should be adopted. 
Actually, once we consider finite $|\vp|$, they yield different results. 

%%%%%%%%%%%%%%%%%%%%%%

%%%%%%%%%%%%%%
\subsubsection{Sum rule 1}

The first sum rule for $\delta\rho_{00}$ is obtained from Eq.~(\ref{eq:derivation}) for $\mu=\nu=0$ by using Eqs.~(\ref{eq:OPE-w-RG-L}) and (\ref{eq:hydro-GR-L}):
\begin{align}
\label{eq:sumrule-1-L}
\begin{split}
\frac{\sigma}{D}
+{\cal O}(\vp^2) 
= \frac{2}{\pi}\int^\infty_{0} d\omega
\frac{\delta \rho_{00}(\omega,\vp)}{\omega}.
\end{split}
\end{align}
We call this sum rule 1 for the longitudinal channel.
Since $\sigma/D$ agrees with the susceptibility $\chi_q$, the left-hand side can be evaluated nonperturbatively by lattice QCD without the problem of analytic continuation.

Let us evaluate the contributions from the diffusion peak, the transport peak, the continuum, and the UV tail, in the weak coupling and the chiral limits.
By using Eq.~(\ref{eq:hydro-rho-L}), the contribution from the diffusion peak is evaluated as
\begin{align}
\begin{split}
\frac{2}{\pi}\int^\infty_{0} d\omega
\frac{\delta \rho_{00}(\omega,\vp)}{\omega}
&= \frac{2}{\pi} \sigma\vp^2
\int^\infty_{0} d\omega \frac{1}{\omega^2+(D\vp^2)^2}\\
&= \frac{\sigma}{D},
\end{split}
\end{align}
which is of order $\Cem T^2$.
This contribution is equal to the left-hand side of Eq.\,(\ref{eq:sumrule-1-L}). 
All the other contributions [the transport peak~(\ref{eq:transport-00}), continuum~(\ref{eq:continuum-00}), 
and the UV tail~(\ref{eq:UVtail-00})] are found to be proportional to $\vp^2$, so the contribution from the diffusion 
peak is dominant when the momentum is small, and sum rule 1 is satisfied in the limit considered here. 

%%%%%%%%%%%%%%
\subsubsection{Sum rule 2}

The second sum rule is obtained by replacing $\delta G^R(\omega)$ [$\delta \rho(\omega)$] with 
$\omega^2\delta G^R(\omega)$ [$\omega^2\delta \rho(\omega)$] in the derivation of sum rule 1, as in Sec.~\ref{sssec:sumrule2-T}.
The result is 
\begin{align}
\label{eq:sumrule-2-L}
0= \frac{2}{\pi}\int^\infty_{0} d\omega \omega
\delta \rho_{00}(\omega,\vp).
\end{align}
This is sum rule 2 for the longitudinal channel, which constrains the spectral function.
We note that this sum rule can be obtained also by using the current conservation~\cite{Bernecker:2011gh}.
Therefore, actually this sum rule is exact, and valid at any value of $|\vp|$, not only for small $|\vp|$.

We again evaluate the contributions to sum rule 2 in the weak coupling and the chiral limits.
First we check the diffusion peak contribution.
By using Eq.~(\ref{eq:hydro-rho-L}), we get the order estimate
\begin{align}
\begin{split}
 \frac{2}{\pi}\int d\omega \omega
\delta \rho_{00}(\omega,\vp) 
& =  \frac{2}{\pi}  \sigma\vp^2 
\int^\infty_{0} d\omega \frac{\omega^2}{\omega^2+(D\vp^2)^2} \\
&\sim \Cem T^2\tau^2 \vp^4 , 
\end{split}
\end{align}
where we have introduced a UV cutoff of the diffusion peak, which is of order $D\vp^2\sim \tau\vp^2$.

The transport peak (\ref{eq:transport-00}) contributes as
\begin{align}
\begin{split}
& \frac{2}{\pi}\int^\infty_{0} d\omega \omega
\delta \rho_{00}(\omega,\vp) \\
&=  \frac{2}{\pi} \frac{2}{3}\Cem\Nc\chi \vp^2
\int^\infty_{0} d\omega 
 \frac{\tau^{-1}}{\omega^2+\tau^{-2}} \\
&~~~\times \left[1
+\vp^2 \frac{2}{5}\left(\tau^{-2}+\frac{11}{3}\omega^2\right)
\frac{1}{(\omega^2+\tau^{-2})^2}
\right] \\
&=   \frac{2}{3}\Cem\Nc\chi \vp^2
 \left[1+\frac{\vp^2  \tau^2}{3} \right] ,
\end{split}
\end{align}
where we have used $\int^\infty_{0} d\omega \tau^{-1}(\tau^{-2}+11\omega^2/3)/(\omega^2+\tau^{-2})^3= \tau^2 (\pi/2)(5/6)$.

The continuum contribution is estimated by using Eq.~(\ref{eq:continuum-00}), 
\begin{align}
\begin{split}
 \frac{2}{\pi}\int^\infty_{0} d\omega \omega
\delta \rho_{00}(\omega,\vp)
&= -C_{\text{em}}\Nc  \frac{\vp^2}{3\pi^2} \int^\infty_{0} d\omega \omega \\
&~~~\times \left[\nf\left(\frac{\omega}{2}\right)+\frac{\vp^2}{40}\nf''\left(\frac{\omega}{2}\right)\right] \\
&= -C_{\text{em}}\Nc  \frac{\vp^2}{3\pi^2}  
 \left[\frac{\pi^2T^2}{3} +\frac{\vp^2}{20}\right], 
\end{split}
\end{align}
where we have used $\int^\infty_{0} d\omega \omega\nf''\left(\omega/2\right)=2$.

The UV tail contribution is estimated to be 
\begin{align}
\begin{split}
 \frac{2}{\pi}\int^\infty_{0} d\omega \omega
\delta \rho_{00}(\omega,\vp)
&= {\cal O}\left( C_{\text{em}} g^2T^2\vp^2, C_{\text{em}} g^2 \vp^4\right),
\end{split}
\end{align}
by using Eq.~(\ref{eq:UVtail-00}).

By looking at all the contributions, we see that the contributions from the transport peak and the continuum are larger than the one from the UV tail at order $\vp^2$. 
However, these two contributions are found to cancel each other, so that sum rule 2 is satisfied at leading order in $\vp^2$. 
For the $\vp^4$ terms, the contributions from the diffusion peak and the transport peak are larger than the ones from the continuum and the UV tail.
These contributions are expected to cancel, but we cannot confirm this here since the former contribution was not explicitly calculated.

%%%%%%%%%%%%%%
\subsubsection{Sum rule 3}

The third sum rule is obtained by replacing $\delta G^R(\omega)$ [$\delta \rho(\omega)$] with 
$\omega^4\delta G^R(\omega)$ [$\omega^4\delta \rho(\omega)$] in the derivation of sum rule 1.
It gives 
\begin{align}
\label{eq:sumrule-3-L}
\begin{split}
&-e^2\sum q^2_f \vp^2 
\Bigl[\left\{2m_f \delta\left\langle\overline{\psi}_f \psi_f \right\rangle
+\frac{1}{12}\delta\left\langle \frac{\alpha_s}{\pi}G^2\right\rangle \right\} \\
&~~~+\frac{8}{3}\frac{1}{4\Cf+\Nf} \delta\left\langle T^{00} \right\rangle  
\Bigr]
= \frac{2}{\pi}\int^\infty_{0} d\omega \omega^3
\delta \rho_{00}(\omega,\vp),
\end{split}
\end{align}
which we call sum rule 3 in the longitudinal channel.
We note that there is no $\vp^4$ correction on the left-hand side, but implicitly the smallness of $|\vp|$ is assumed so that hydrodynamics is reliable. 
As before, the operators on the left-hand side of this sum rule can be evaluated by lattice QCD without having the problem of analytic continuation.

Let us evaluate the contribution to this sum rule in the weak coupling and the chiral limits.
The diffusion peak contribution is, by using Eq.~(\ref{eq:hydro-rho-L}), estimated as 
\begin{align}
\begin{split}
 \frac{2}{\pi}\int^\infty_{0} d\omega \omega^3
\delta \rho_{00}(\omega,\vp)
&= \frac{2}{\pi}\sigma\vp^2
\int^\infty_{0} d\omega 
\frac{\omega^4}{\omega^2+(D\vp^2)^2}\\
&\sim \Cem T^2 \tau^4\vp^8,
\end{split}
\end{align}
where we have used the fact that the UV cutoff of the diffusion peak is of order $\tau \vp^2$.

The transport peak contributes as 
\begin{align}
\begin{split}
& \frac{2}{\pi}\int^\infty_{0} d\omega \omega^3
\delta \rho_{00}(\omega,\vp) \\
&=  \frac{2}{\pi}\frac{2}{3}\Cem\Nc\chi \vp^2 
\int^\infty_{0} d\omega  \frac{\omega^2 \tau^{-1}}{\omega^2+\tau^{-2}} \\
&~~~\times \left[1
+ \frac{2}{5}\left(\tau^{-2}+\frac{11}{3}\omega^2\right)
\frac{\vp^2}{(\omega^2+\tau^{-2})^2} \right] \\
&= {\cal O}(\Cem T^2\vp^2\tau^{-1}\varLambda)
+ \Cem\Nc\chi \vp^4 \frac{2}{5} ,
\end{split}
\end{align}
where we have used $\int^\infty_{0} d\omega \left(\tau^{-2}+11\omega^2/3\right) \omega^2 \tau^{-1}/(\omega^2+\tau^{-2})^3=(\pi/2)(3/2)$.

The contribution from the continuum (\ref{eq:continuum-00}) reads
\begin{align}
\begin{split}
 \frac{2}{\pi}\int^\infty_{0} d\omega \omega^3
\delta \rho_{00}(\omega,\vp)
&= -C_{\text{em}}\Nc  \frac{\vp^2}{6\pi}  \frac{2}{\pi}\int^\infty_{0} d\omega \omega^3 \\
&~~~\times \left[\nf\left(\frac{\omega}{2}\right)+\frac{\vp^2}{40}\nf''\left(\frac{\omega}{2}\right)\right] \\
&= -C_{\text{em}}\Nc  \frac{\vp^2}{3}  \\
&~~~\times \left[\frac{14\pi^2T^4}{15}+ \frac{\vp^2}{5}T^2\right]. 
\end{split}
\end{align}

Finally, the UV tail contribution (\ref{eq:UVtail-00}) is 
\begin{align}
\begin{split}
 \frac{2}{\pi}\int^\infty_{0} d\omega \omega^3
\delta \rho_{00}(\omega,\vp)
&= \frac{2}{\pi}\Cem\alpha_s(\kappa_0) \Nc\Cf \frac{4\pi^2T^4}{27} \vp^2 \\
&~~~\times \int^\infty_{X_0} dX
  \left[\frac{X_0}{X}\right]^{\tilde{a}+1} \\ 
&~~~+{\cal O}(\Cem g^2T^2\vp^4)\\
  &= \Cem \Nc \vp^2 \frac{8\pi^2T^4}{9} 
 \frac{\Cf}{4\Cf+\Nf} \\
&~~~ +{\cal O}(\Cem g^2T^2\vp^4).
\end{split}
\end{align}

Let us check the sum rule order by order in $\vp^2$. 
For the $\vp^2$ terms, we see that the contributions from the continuum and the UV tail are dominating, 
and their sum is equal to $-C_{\text{em}}\Nc\vp^2\pi^2T^4 2(8\Cf+7\Nf)/[45(4\Cf+\Nf)]$, which agrees with the left-hand side of sum rule 3 (\ref{eq:sumrule-3-L}).
The order $\vp^4$ terms are dominated by the transport peak and the continuum, but they cancel each other. 
Therefore, sum rule 3 is shown to be satisfied up to order $\vp^4$. 

%%%%%%%%%%%%%%
\subsubsection{Sum rules for $\delta\rho_L$}

Before ending this section, we note that two sum rules for $\delta\rho_L$ can be derived from our three sum rules for $\delta\rho_{00}$, 
(\ref{eq:sumrule-1-L}), (\ref{eq:sumrule-2-L}), and (\ref{eq:sumrule-3-L}). 
Using $\rho_L(p)=p^2 \rho_{00}(p)/\vp^2$, we get
\begin{align}
& -\frac{\sigma}{D}
+{\cal O}(\vp^2) 
= \frac{2}{\pi}\int^\infty_{0} d\omega
\frac{\delta \rho_{L}(\omega,\vp)}{\omega}, \\
\nonumber
& -e^2\sum q^2_f  
\Bigl[\left\{ 2m_f  \delta\left\langle\overline{\psi}_f \psi_f \right\rangle
+\frac{1}{12}\delta\left\langle \frac{\alpha_s}{\pi}G^2\right\rangle \right\} \\
&+\frac{8}{3}\frac{1}{4\Cf+\Nf} \delta\left\langle T^{00} \right\rangle  
\Bigr] 
= \frac{2}{\pi}\int^\infty_{0} d\omega \omega
\delta \rho_{L}(\omega,\vp)
\end{align}
Naively, one would expect these to agree at $|\vp|=0$
with the corresponding sum rules for $\delta \rho_T$, Eqs.\,(\ref{eq:sumrule-1-T}) and (\ref{eq:sumrule-2-T}). 
This is indeed the case for the latter sum rule, but the former one does not agree, because of the $-\sigma/D$ term on its left-hand side. 
This disagreement can be traced back to the singularity of $\delta \rho_L$ at the diffusion pole $\omega=-iD\vp^2$. 

%%%%%%%%%%%%%%%%%%%%%%%%%%%%%%%%%%%%%%%%%%%
\section{Application to lattice QCD data analysis}
\label{sec:lattice}

In this section, we demonstrate that our sum rules can be used to improve a fit of the spectral function to the latest Euclidean time lattice QCD data given in Ref.\,\cite{Brandt:2015aqk}.
In doing this, we make use of the sum rules in three different ways. 
\begin{enumerate}
\item Providing guidance in the choice of the functional forms used to parametrize the 
spectral function. 
\item Reducing the number of fitting parameters (sum rules 1 and 2). 
\item Determining transport coefficients (sum rule 3). 
\end{enumerate}
We note that we use all the sum rules including sum rule 2, which was missing in the previous work~\cite{Gubler:2016hnf}.

%In the previous sections, we have derived and checked several sum rules that are applicable to the longitudinal and transverse components of the spectral function at finite momentum. 
In the following discussion of lattice QCD data, we restrict ourselves to the zero-momentum case, 
as most currently available lattice QCD data are provided only in this limit. 
This simplifies the situation in the sense that for $|\vp| = 0$ the correlator has only one independent component and 
some parameters drop out of the sum rules [such as $\kappa_B$ in Eq.\,(\ref{eq:sumrule-1-T})]. 
In this section, we use the notation $\rho(\omega, T) = \rho_{T}(\omega, |\vp|=0) = \rho_{L}(\omega, |\vp|=0)$ for the spectral function. 
The application of our sum rules to nonzero momentum lattice QCD data is left for future work. 
We hence only use the sum rules of Eqs.\,(\ref{eq:sumrule-1-p0}) and (\ref{eq:sumrule-2-T}) 
at $|\vp| = 0$ and (\ref{eq:sumrule-3-T}). 

\subsection{Parametrization of the spectral function in vacuum and at finite temperature}
Our goal here is to find a functional form of the spectral function that is consistent with the sum rules 1-3, 
that is, that does not lead to divergent results for these sum rules. 
Note that some terms of the finite temperature spectra used in Ref.\,\cite{Brandt:2015aqk} in fact lead to divergences for sum rules 2 and 3 and therefore violate them. 

\begin{figure}
\begin{center}
%\vspace{-1.0cm}
%\hspace*{1.2cm}
\includegraphics[width=0.49\textwidth]{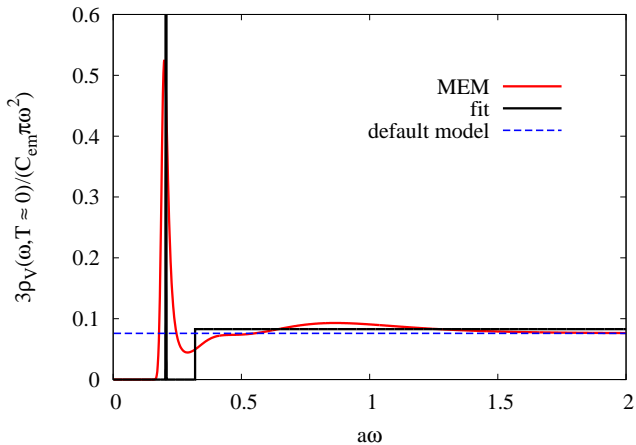}
%\vspace{0.5cm}
\caption{Result of a MEM analysis of vacuum data of Ref.\,\cite{Brandt:2015aqk} (red solid line), compared 
with a simple fit using Eq.\,(\ref{eq:vac}) (black solid line). The blue dashed line shows the default model 
used in the MEM analysis.}  
\label{fig:MEM.fit}
\end{center} 
\end{figure}

First, we start with the spectral function in vacuum, for which we can follow the parametrization of  
Ref.\,\cite{Brandt:2015aqk}, 
\begin{align}
& \frac{1}{C_{\mathrm{em}}} \rho(\omega, T \simeq  0) \nonumber \\
=& \,\, \frac{\pi}{3} a_V \delta(\omega - m_V)  \nonumber \\
&\,\, + (1 + k_1) \frac{1}{4 \pi} \Theta(\omega - \Omega_0) 
                                                \omega^2 \tanh \Bigl( \frac{\omega \beta_0}{4} \Bigr)
\label{eq:vac}                                                
\end{align}
This form is adapted to our notation, which differs from that of Ref.\,\cite{Brandt:2015aqk} by 
a factor of $1/6$ and the treatment of $C_{\mathrm{em}}$. 
Here $\beta_0 = 1/T_0$, where $T_0$ corresponds to the low temperature of the ``vacuum" lattice ensemble 
of  Ref.\,\cite{Brandt:2015aqk} ($T_0 \simeq 32\,\mathrm{MeV}$). The values of the parameters $a_V$, $m_V$, $k$ 
and $\Omega_0$ are determined by the fit. Here, the $\delta$-function peak corresponds to the $\rho$ meson, 
as an isospin 1 current was used in Ref.\,\cite{Brandt:2015aqk}. To check whether the above is a reasonable 
parametrization, we have performed a MEM analysis~\cite{Jarrell:1996rrw,Asakawa:2000tr,Gubler:2010cf} 
of the vacuum data provided in Ref.\,\cite{Brandt:2015aqk}. 
In this simple analysis, we have ignored correlations between data points at different time slices. The result 
is shown as a red line in Fig.\,\ref{fig:MEM.fit}. Here, we have chosen the default model, which is an input of the MEM algorithm, 
to match the perturbative value of the spectral function at high energy [$\rho(\omega)/\omega^2 =  3/(2\pi)$, blue dashed line in Fig.\,\ref{fig:MEM.fit}]. 
It is seen in the figure that at low energy the spectral function is dominated by a single peak, while at high energy, there 
is an almost flat continuum. We furthermore see hints of excited states at around $a\omega \simeq 0.8$, but their effect seems to 
to be weak. 
In all, Eq.\,(\ref{eq:vac}) turns out to be a reasonable, if somewhat rough, parametrization of the spectral function 
at temperatures much below $T_c$. 
As is also shown in Fig.\,\ref{fig:MEM.fit} and further discussed later, 
the fit results for the peak position $m_V$ and the onset of the continuum  $\Omega_0$ 
agree well with those obtained from the MEM analysis. 

Next, we consider the spectral function at temperatures around and above $T_c$, where large modifications 
are expected. As it is already discussed in detail in this work, we most importantly expect the appearance of a transport peak 
at $\omega \simeq 0$ and an UV tail at high energy. We again follow the parametrization used in Ref.\,\cite{Brandt:2015aqk}, 
but modify it such that sum rules 1-3 can be satisfied. Specifically, we use 
\begin{align}
& \frac{1}{C_{\mathrm{em}}} \rho(\omega, T) \nonumber \\
=&\,\, \frac{\omega A_T \Gamma_T}{3(\Gamma_T^2 + \omega^2)} \bigl[ 1 - A(\omega) \bigr]
                                + \frac{\pi}{3} a_T \delta(\omega - m_V) \nonumber \\ 
                                 &\,\, + A(\omega) \bigl[1 + \tilde{k}(\omega) \bigr] \frac{1}{4 \pi} 
                                 \omega^2 \tanh \Bigl( \frac{\omega \beta_T}{4} \Bigr) \nonumber \\
                                &\,\, + \frac{c_0}{4 \pi} \Theta(\omega - \Omega_0) \frac{1}{\omega^2} \frac{1}{\bigl[\ln(\omega/\Lambda_{\mathrm{QCD}})\bigr]^{1+ \tilde{a}}},  
\label{eq:finiteT}                                                
\end{align}
with
\begin{align}
A(\omega) = \tanh\Bigl( \frac{\omega^2}{\Delta^2}\Bigr) 
\label{eq:A}
\end{align}
and
\begin{align}
\tilde{k}(\omega) = k_1 + k_2 \Biggl[1 - \tanh^2\Bigl( \frac{\omega}{\Omega_0 \eta} \Bigr) \Biggr]. 
\label{eq:k}
\end{align}
Let us here mention the differences between our above parametrization and that of Ref.\,\cite{Brandt:2015aqk}. 
First, the factors $\bigl[ 1 - A(\omega) \bigr]$ and 
$A(\omega)$, which were introduced already in our previous work~\cite{Gubler:2016hnf}, are there for cutting off the divergence in the integral of sum rules 2 and 3.
Furthermore, the factor $1/\bigl[\ln(\omega/\Lambda_{\mathrm{QCD}})\bigr]^{1+ \tilde{a}}$, which is adapted from the perturbative expression, 
is employed to make integral of sum rule 2 finite.

\subsection{Lattice data used in the fit}
Reference~\cite{Brandt:2015aqk} provides correlator data of altogether five lattice ensembles with respective temperatures 
$32$, $169$, $203$, $254$ and $338\,\mathrm{MeV}$, which 
in the setting of that work corresponds to $0.15$, $0.8$, $1.0$, $1.25$ and $1.76\,T_c$. The spatial extent of the 
lattice is $N_s = 64$ for all ensembles, while the temporal size is $N_{\tau} = 128$, 24, 20, 16 and 12. The lattice spacing is 
$a = 0.0486\,\mathrm{fm}$ and the vacuum pion mass $270\,\mathrm{MeV}$, which shows that this is not yet a physical point 
simulation. For more details, we refer the reader to the original publication of Ref.\,\cite{Brandt:2015aqk}. 

The lattice data are related to the spectral function $\rho(\omega)$ through the following integral: 
\begin{align}
G^{\mathrm{E}}(\tau, T) = \int_{0}^{\infty} \frac{d \omega}{2 \pi} \frac{\rho(\omega,T)}{C_{\mathrm{em}}} \frac{\cosh[\omega(\tau - 1/2T)]}{\sinh(\omega/2T)}.
\label{eq:Eucl.vec.correlator}
\end{align} 
Here, $\tau$ is defined in the interval $0 \leq \tau < 1/T$ and is symmetric with respect to the central point $\tau = 1/(2T)$. Hence, 
only half of the temporal data points can be used as independent information for the fit. Furthermore, for small $\tau$ values, the data 
could contain lattice artifacts. We follow  Ref.\,\cite{Brandt:2015aqk} and use data points $\tau/a \in [4:48]$ for the vacuum 
ensemble and $\tau/a \in [4:N_{\tau}/2]$ for the others. 

\subsection{Treatment of the parameters}
We treat the various parameters introduced in Eqs.\,(\ref{eq:vac})-(\ref{eq:k}) as follows. We keep $m_V$, $k_1$ and $\Omega_0$ 
fixed for both vacuum and finite temperature. $\eta$, which only enters in the nonzero temperature parametrization, 
is also kept fixed for all ensembles. All other parameters, $A_T$, $\Gamma_T$, $\Delta$, $a_T$, $k_2$ and $c_0$ are 
allowed to depend on temperature. To reduce the numbers of unknowns to be fitted, we however make use of the sum rules 1 and 2 
to constrain the parameters. To be specific, the sum rules are used to fix $A_T/\Gamma_T$ and $c_0$ at each temperature. 
To be able to use sum rule 2, one, in principle, needs the quark- and gluon-condensate values and the energy density as a function 
of temperature. For the simple trial analysis of this work, we employ the values provided in Ref.\,\cite{Bazavov:2014pvz}, which we rescale to two flavors 
according to the ratio in the perturbative high-temperature limit. As the quark masses and other lattice parameters in Ref.\,\cite{Bazavov:2014pvz} are 
different from those of  Ref.\,\cite{Brandt:2015aqk}, this is not a completely consistent treatment and will have to be improved in a more 
quantitative analysis in the future. 

\subsection{Fit results}
Performing our fit, it turns out that even if one uses both sum rules 1 and 2 as constraints, the lattice data are not sufficient to determine 
all parameters of Eqs.(\ref{eq:finiteT})-(\ref{eq:k}) with good accuracy. We found for instance that certain combinations of 
the ratio $A_T/\Gamma_T$, which is proportional to 
the electric conductivity $\sigma_{\mathrm{el}}$, and $\Delta$ lead to similar $\chi^2$ values. This means that the systematic uncertainty of these quantities 
is large and that more data points with smaller errors will be needed to further constrain the parameters, especially for the lattice 
ensembles with temperatures above $T_c$, where the number of usable data points is rather small.  
In the following we provide one possible fit result, which is obtained by demanding that $\sigma_{\mathrm{el}}/T$ is a monotonically increasing 
function of $T$ and take values in reasonable agreement with previous works \cite{Brandt:2015aqk,Aarts:2007wj,Francis:2011bt,Ding:2010ga,
Ding:2013qw,Ding:2014dua,Amato:2013naa,Brandt:2012jc}. 
The numerical result of the fit is given in Table\,\ref{tab:fit.values}. These give a $\chi^2$ value of 
$\chi^2/\mathrm{d.o.f.} = 0.87$. 
We emphasize once more that this is not 
the only solution and should therefore be regarded as an illustrative example rather than the final result.  
Nevertheless, the purpose of our analysis is to demonstrate the possibility to use the sum rules to improve the lattice QCD analysis, 
and to this end, we believe that our fit suffices.

\begin{table*}
\renewcommand{\arraystretch}{1.3}
\begin{center}
\caption{Parameter values obtained by our fit using lattice QCD data. The dimensionfull parameters describing the zero-temperature spectral function ($m_V$, $a_V$, $\Omega_0$) 
are given in lattice units ($k_1$ is dimensionless). All other parameters are made dimensionless by dividing them by appropriate powers of $T$.} 
\label{tab:fit.values}
\begin{tabular}{cccccc}  
\hline 
Fitted parameters & Fixed & $0.8\,T_c$ & $1.0\,T_c$ & $1.25\,T_c$ & $1.67\,T_c$ \\ \hline
$m_V$ & 0.206 & - & - & - & - \\
$a_V$ & 0.000860 & - & - & - & - \\
$\Omega_0$ & 0.319 & - & - & - & - \\
$k_1$ & 0.0909 & - & - & - & - \\
$\eta$ & 2.04 & - & - & - & - \\
$a_T/T^3$ & - & 4.15 & 2.94 & 2.15 & 1.08 \\
$\Gamma_T/T$ & - & 1.36 & 0.54 & $\infty$ & $\infty$ \\
$A_T/(\Gamma_T T)$ & - & 0.450 & 0.728 & 0.728 & 0.758 \\
$\Delta_T/T$ & - & 4.06 & 2.97 & 0.79 & 1.23 \\
$c_0/T^3$ & - & -166 & -250 & -186 & - 95 \\
$k_2$ & - & -0.0294 & -0.0377 & -0.0322 & -0.0040 \\
\hline
\end{tabular}
\end{center}
\end{table*}

\begin{figure}
\begin{center}
%\vspace{-1.0cm}
%\hspace*{1.2cm}
\includegraphics[width=0.49\textwidth]{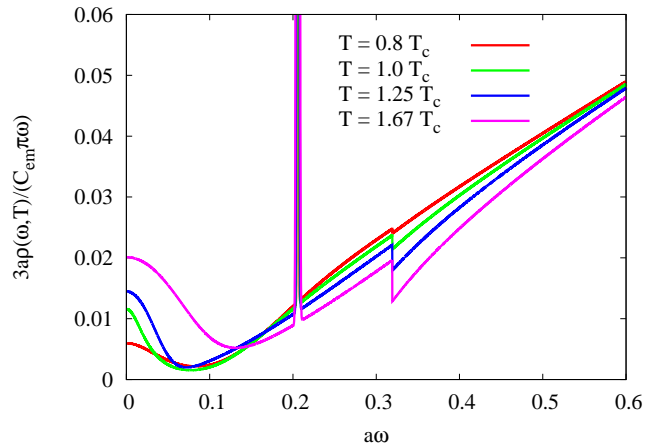}
%\vspace{0.5cm}
\caption{The fitted spectral functions for temperatures between $0.8$ and $1.67\,T_c$. To improve the visibility, 
we have transformed the $\rho$ meson peak, which in the fit is assumed to be a delta function, into a Gaussian with a small, 
but nonzero width.}  
\label{fig:full.spec}
\end{center} 
\end{figure}

Let us now discuss the resulting spectral functions in some detail. First, we compare in Fig.\,\ref{fig:MEM.fit}  the vacuum spectral function of 
Eq.\,(\ref{eq:vac}) with that obtained by MEM. 
It is seen that the two methods indeed give a quantitatively similar result, while the details naturally disagree 
due to the roughness of the parametrization of Eq.\,(\ref{eq:vac}) and the limited resolution of MEM. 
Next, we consider the spectral functions for temperatures around and above $T_c$. The spectra are shown 
in Fig.\,\ref{fig:full.spec}. 
It is seen in this figure that the transport peak gradually grows with increasing temperature. 
One furthermore observes 
from the behavior of the discontinuity around $a\omega\simeq 0.3$
that the UV tail, which is parametrized to appear above $\Omega_0$, grows with larger temperature. 
We note that the coefficient of the UV tail ($c_0$) is negative in this fit, which might indicate that $\langle \tilde{T}^{00}(\omega\sim T)\rangle $ is negative, 
as is suggested from Eq.~(\ref{eq:deltarho_T_UVtail}).
Also note that we here have divided the spectral function by $\omega$, 
which means that it approaches a constant at small $\omega$, while it approaches a form proportional to $\omega$ at large energy. 

\begin{figure}
\begin{center}
%\vspace{-1.0cm}
%\hspace*{1.2cm}
\includegraphics[width=0.49\textwidth]{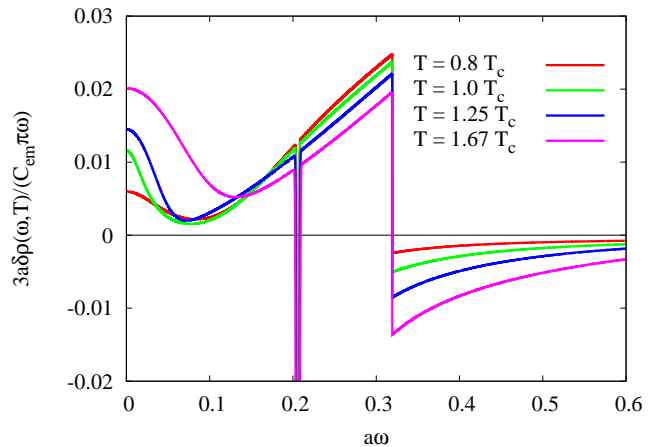}
%\vspace{0.5cm}
\caption{Same as Fig.\,\ref{fig:full.spec}, but with vacuum part subtracted.}  
\label{fig:subtr.spec}
\end{center} 
\end{figure}

To examine the behavior of the spectral function more in detail, we show the same spectral functions with the 
vacuum part subtracted in Fig.\,\ref{fig:subtr.spec}. 
At low energy, the same features can be observed as in the previous figure. At high energy, the dominant part 
of the continuum gets subtracted and only the UV tail remains. 
Also note that because the residue of the $\rho$ meson mass is reduced at finite temperature, the residue 
becomes negative for the subtracted spectral function shown here. 

\begin{figure*}
\begin{center}
%\vspace{-1.0cm}
%\hspace*{1.5cm}
\includegraphics[width=0.45\textwidth]{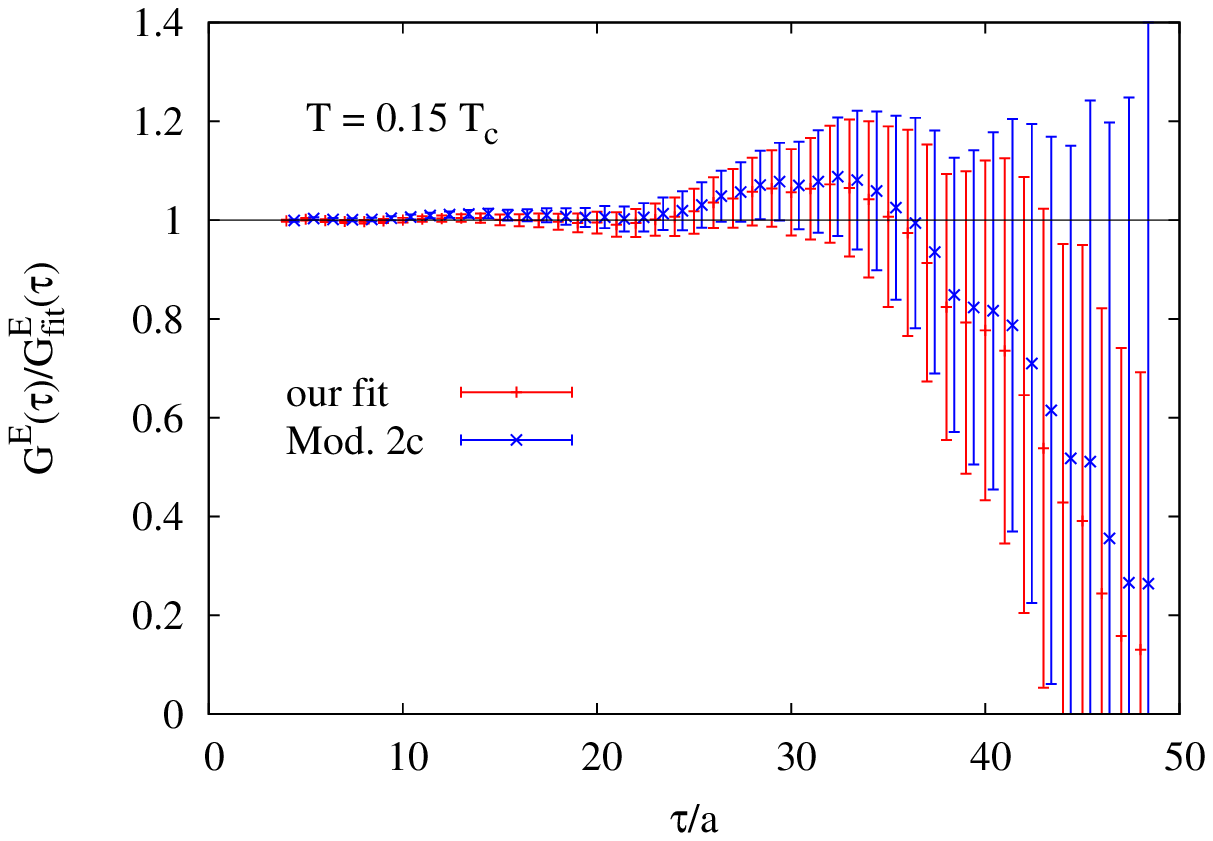}
\includegraphics[width=0.45\textwidth]{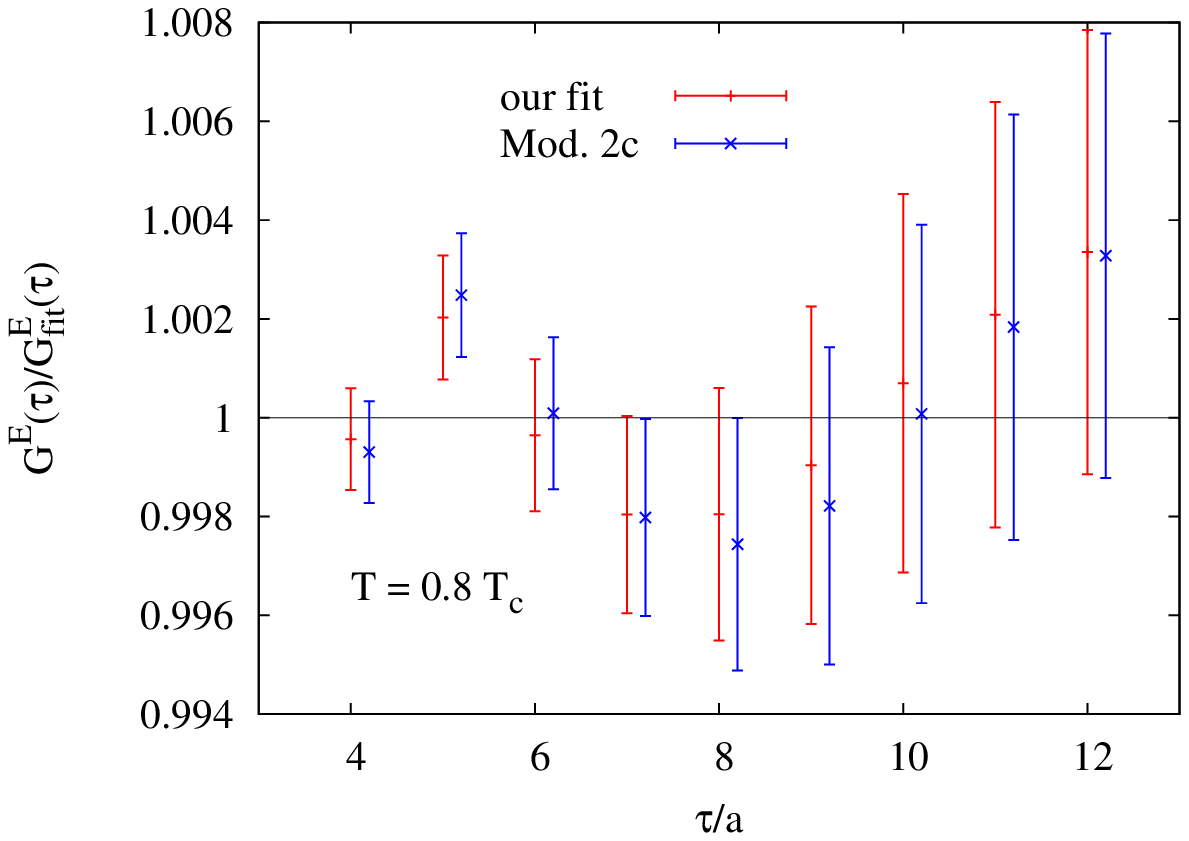}
%\hspace*{1.5cm}
\includegraphics[width=0.45\textwidth]{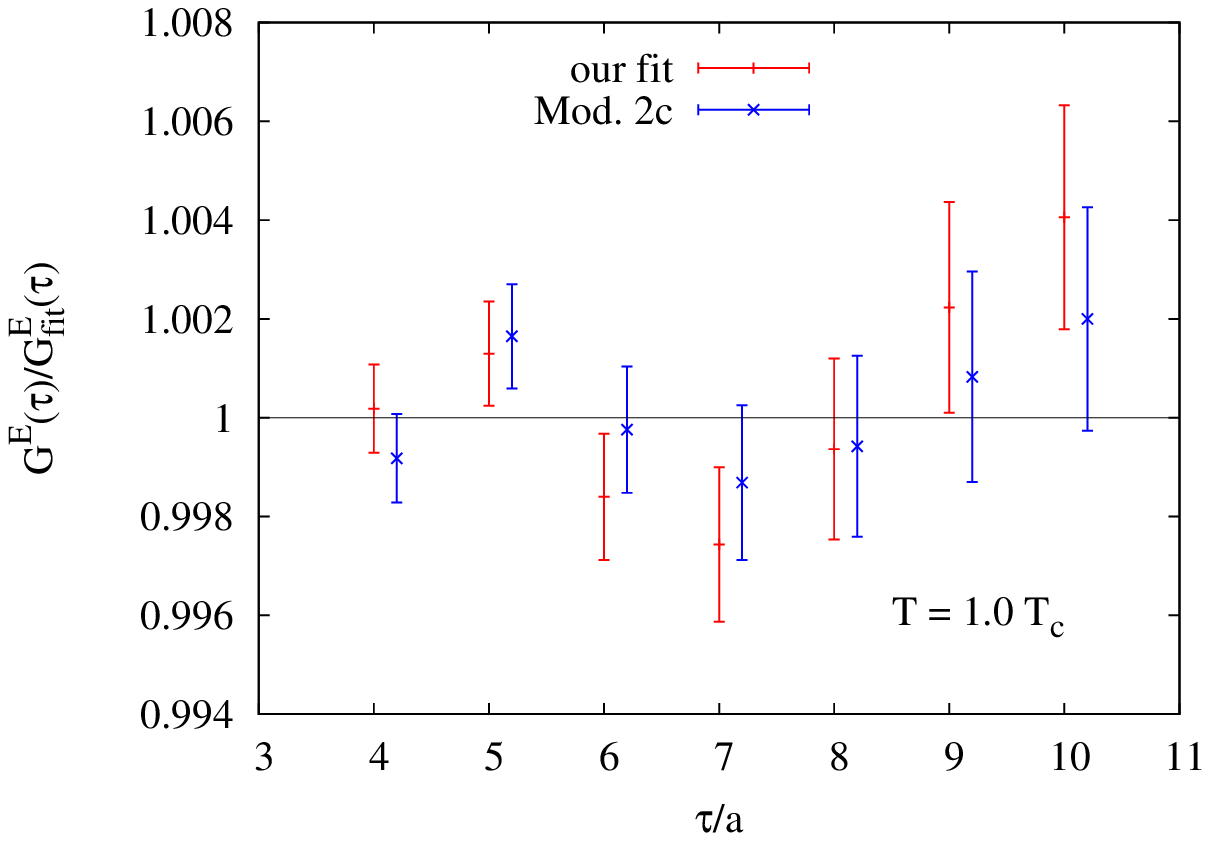}
\includegraphics[width=0.45\textwidth]{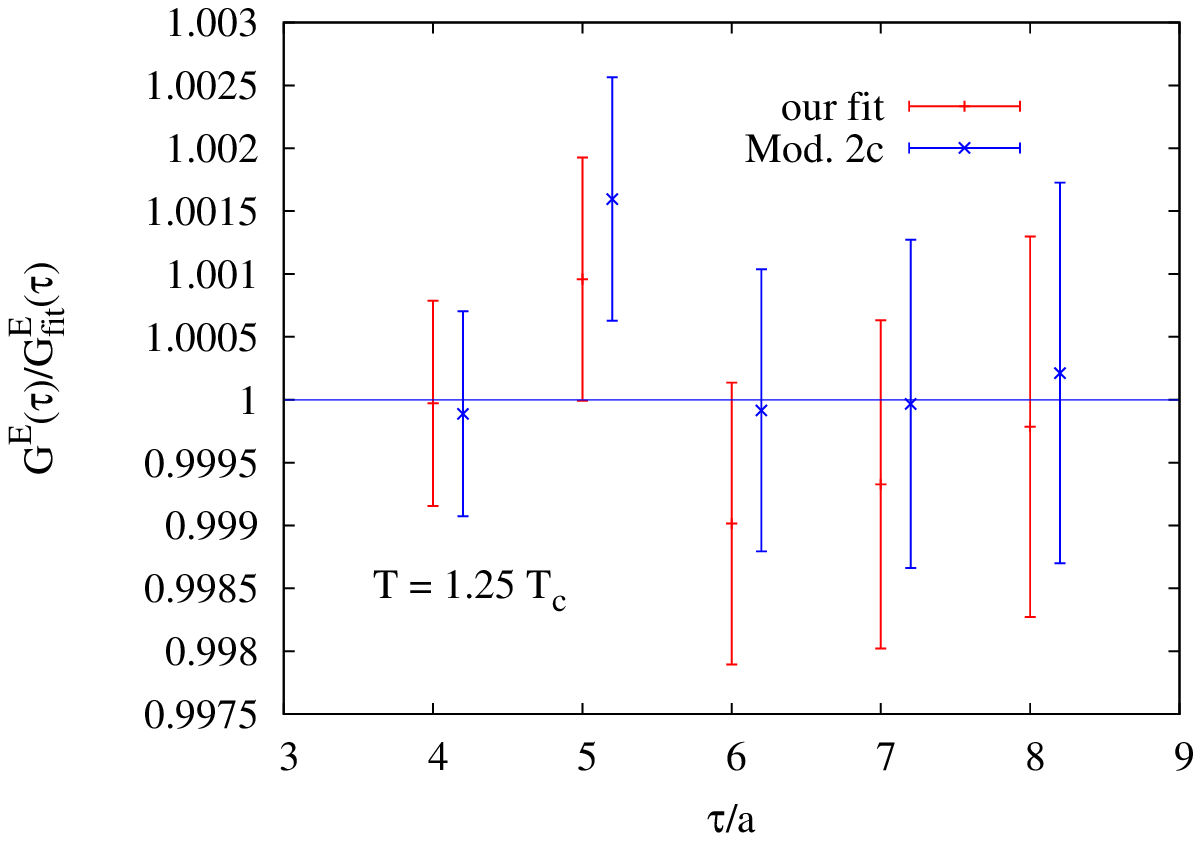}
%\hspace*{1.5cm}
\includegraphics[width=0.45\textwidth]{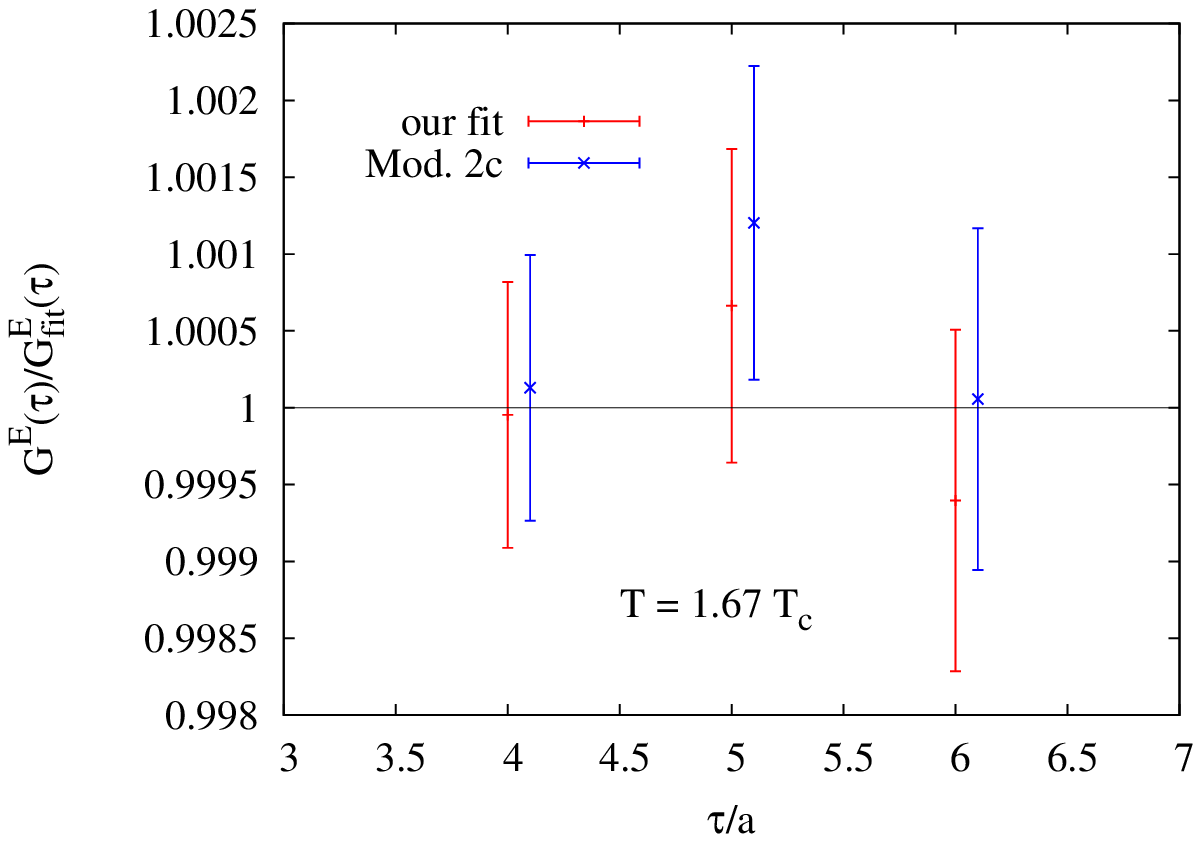}
%\vspace{1.0cm}
\caption{
Ratio of the lattice data of Ref.\,\cite{Brandt:2015aqk}, $G^{\mathrm{E}}(\tau)$, 
and the numerical integral of the fitted spectral function $G^{\mathrm{E}}_{\mathrm{fit}}(\tau)$ [right-hand side of Eq.\,(\ref{eq:Eucl.vec.correlator})] 
at $T = 0.15$, $0.8$, $1.0$, $1.25$ and $1.67\,T_c$. 
Here only the data points that were used in the fit are shown. 
The red data points show the ratio for our fitted spectral function, while one representative result of the fits performed in Ref.\,\cite{Brandt:2015aqk} 
(model 2c) is shown as blue data points. 
For better visibility, the blue data points of model 2c are slightly shifted to the right.}  
\label{fig:comp.latt}
\end{center} 
\end{figure*}

Next, let us examine the quality of our fit by numerically integrating the spectral function as in 
Eq.\,(\ref{eq:Eucl.vec.correlator}) and compare the result with the lattice data of Ref.\,\cite{Brandt:2015aqk}. 
This comparison is shown as red and blue data points in Fig.\,\ref{fig:comp.latt}. 
In these plots we show the ratio between the lattice data and the 
integrated spectral function, which should be 1 for a perfect fit. Overall, it 
can be seen that both our and the fit of Ref.\,\cite{Brandt:2015aqk} can reproduce the 
lattice data fairly well, especially if one considers the very small errors of most of the 
lattice data points. 
For the lowest temperature ($T = 0.15\,T_c$), both fits 
exhibit almost the same behavior for small $\tau/a$ values, while for $\tau/a \gtrsim 40$, the fit 
of Ref.\,\cite{Brandt:2015aqk} works slightly better than ours. 
At higher temperatures, both fits perform equally well, being consistent 
with the lattice results within errors for most $\tau/a$ values. 
This shows that the available lattice 
data are not sufficient to distinguish our finite temperature spectral functions from those 
obtained in Ref.\,\cite{Brandt:2015aqk}.

\subsection{Comparison with fit of Ref.\,\cite{Brandt:2015aqk}}
In Ref.\,\cite{Brandt:2015aqk} a fit similar to ours was performed with the following functional form. For the 
vacuum part, the same parametrization as in Eq.\,(\ref{eq:vac}) was employed, while for nonzero temperature 
\begin{align}
& \frac{1}{C_{\mathrm{em}}} \rho(\omega, T) \nonumber \\
&= \,\, \frac{\omega A_T \Gamma_T}{3(\Gamma_T^2 + \omega^2)} 
                                + \frac{\pi}{3} a_T \delta(\omega - m_V)  \nonumber \\
&~~~ + \Theta(\omega - \Omega_T) \bigl[1 + \tilde{k}(\omega) \bigr] \frac{1}{4 \pi} 
                                 \omega^2 \tanh \Bigl( \frac{\omega \beta_T}{4} \Bigr) \nonumber \\
&~~~ + \frac{c_0}{4 \pi} \Theta(\omega - \Omega_0) \frac{1}{\omega^2} 
\label{eq:finiteT.Brandt}                                                
\end{align}
was used, which we have again adapted to our notation. 
It should be noted that this form is contradicting with the sum rule 2, as its first and fourth terms lead to divergences 
in that sum rule. If one however only uses sum rule 1, a fit is possible. The authors of Ref.\,\cite{Brandt:2015aqk} have tested several models, 
in which some combinations 
of parameters are set to 0 or other fixed values for certain temperatures. For details, we refer the reader to Ref.\,\cite{Brandt:2015aqk}. 

\begin{figure*}
\begin{center}
%\vspace{-1.0cm}
%\hspace*{1.5cm}
\includegraphics[width=0.45\textwidth]{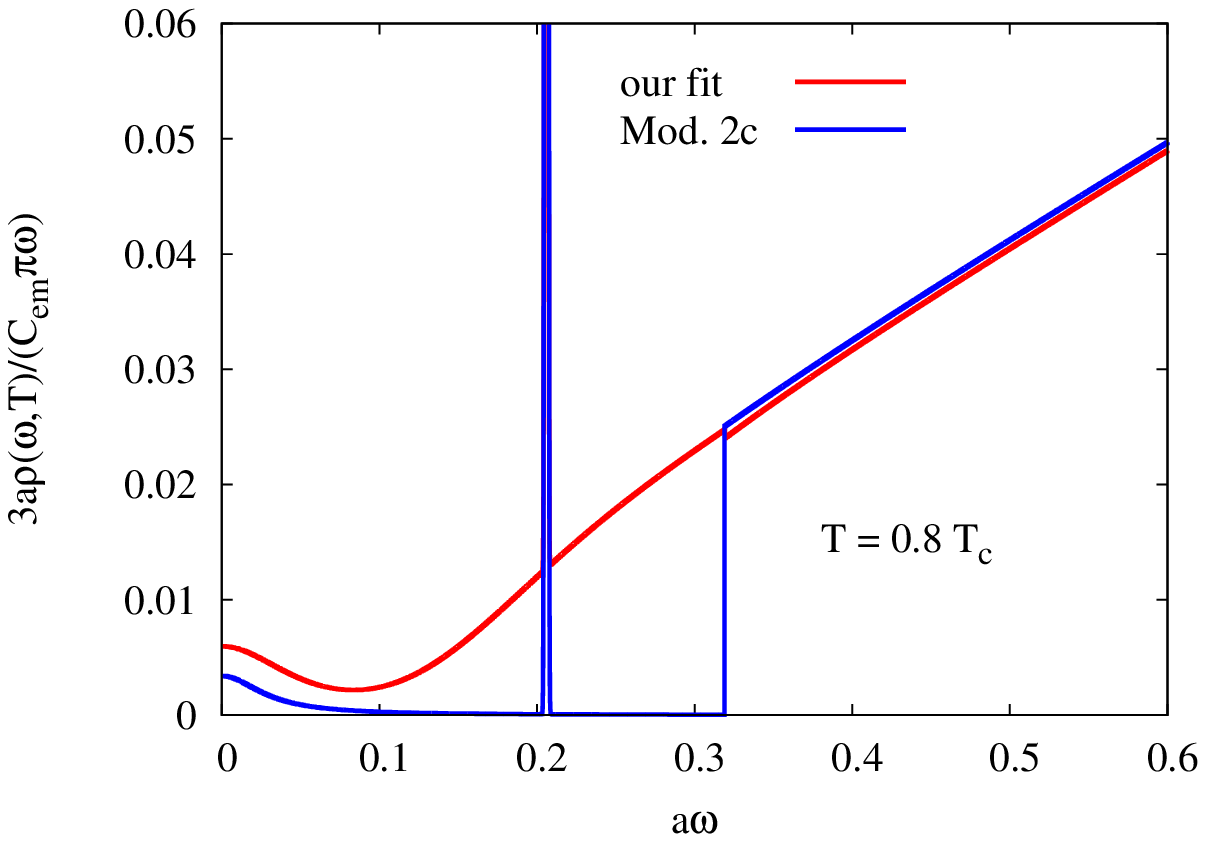}
\includegraphics[width=0.45\textwidth]{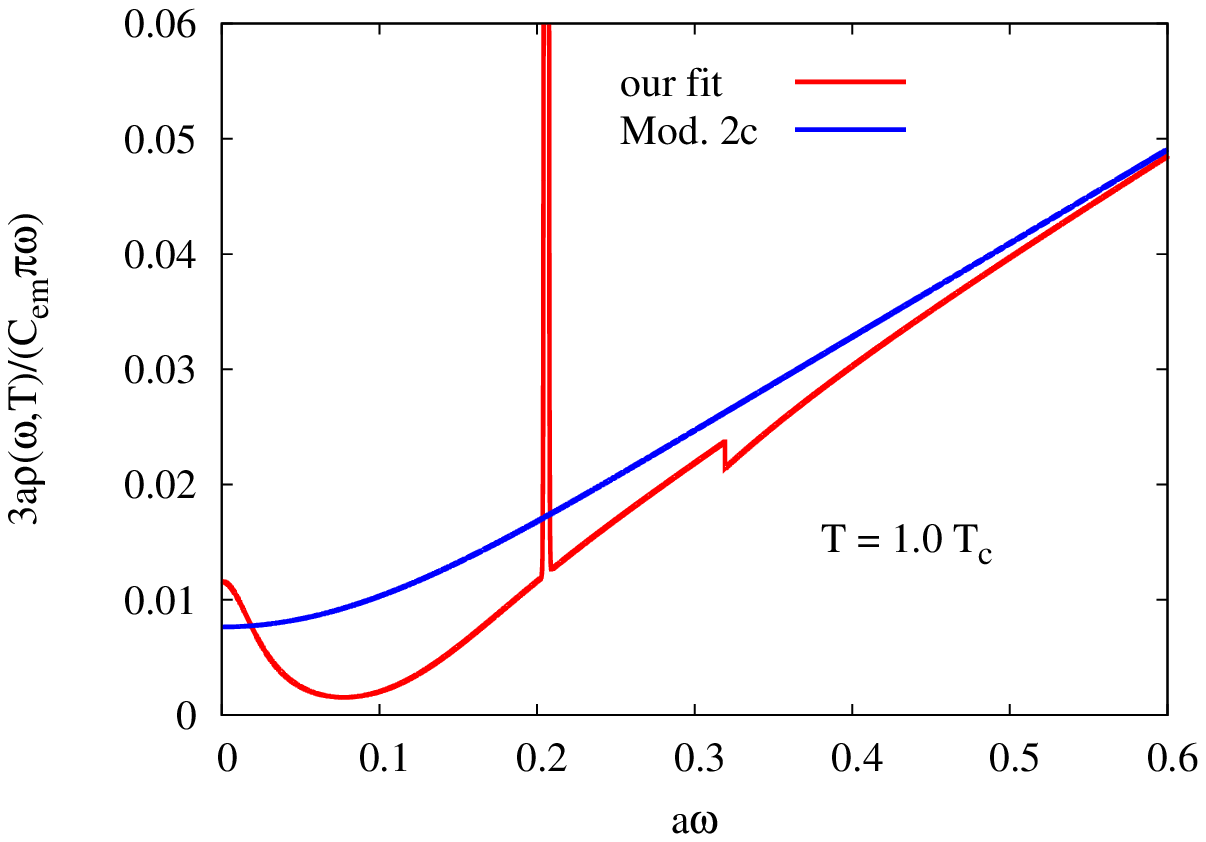}
%\hspace*{1.5cm}
\includegraphics[width=0.45\textwidth]{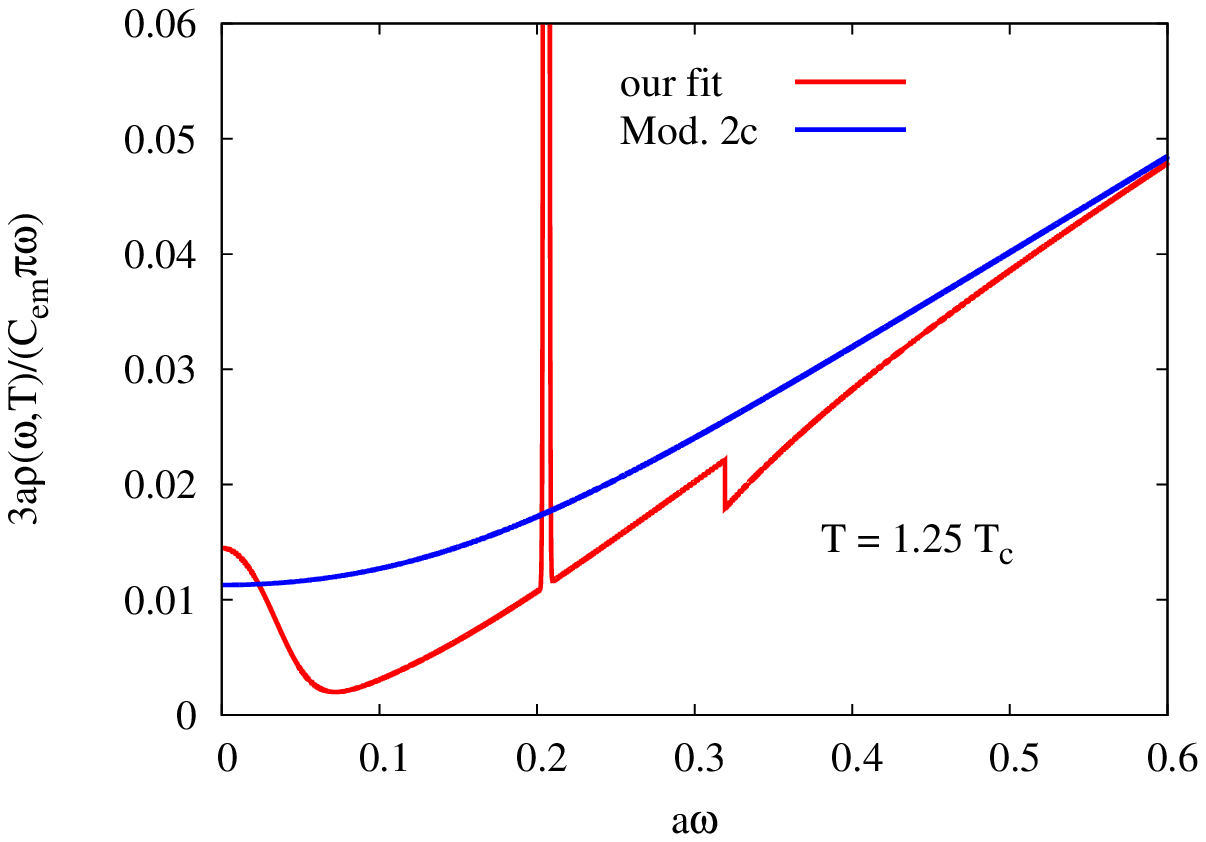}
\includegraphics[width=0.45\textwidth]{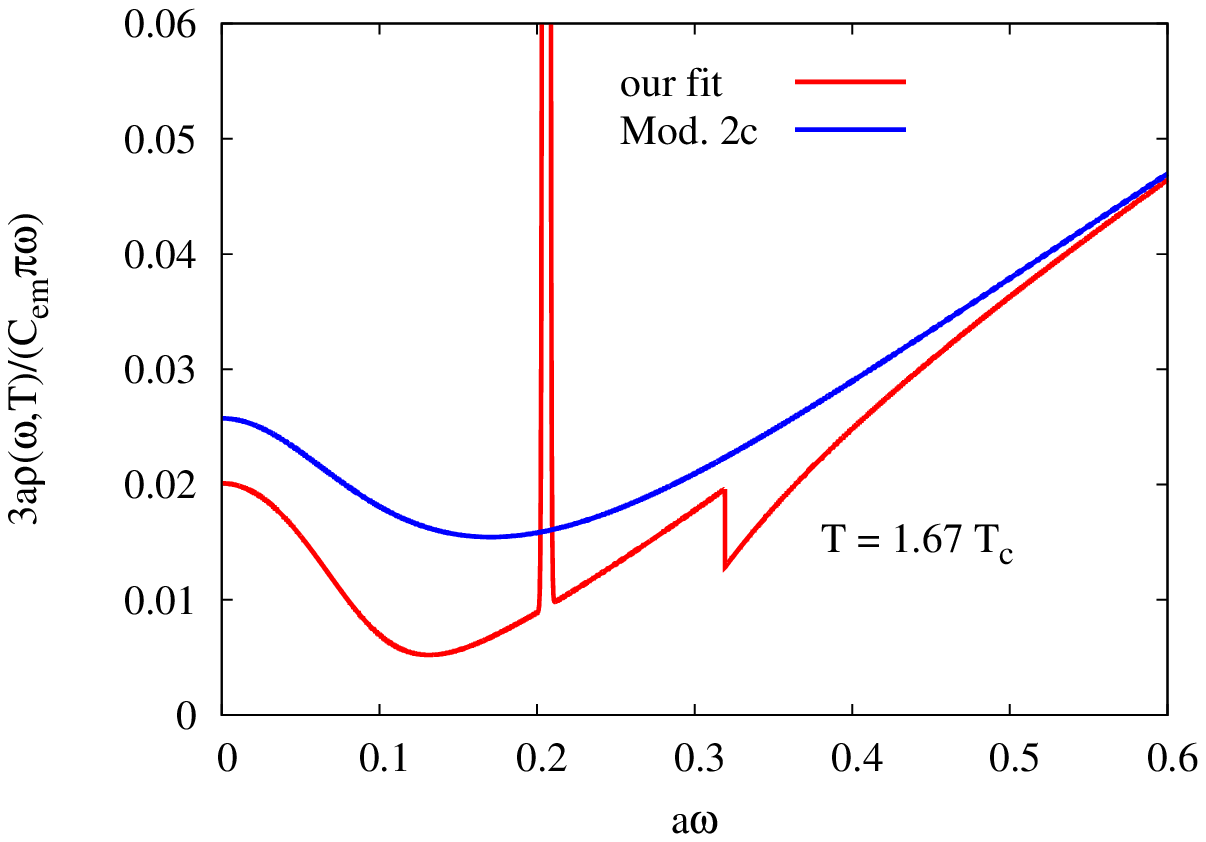}
%\vspace{1.0cm}
\caption{The full spectral functions obtained from our fit (red lines) compared to Model 2c of Ref.\,\cite{Brandt:2015aqk} (blue lines).}  
\label{fig:full.spec.comp}
\end{center} 
\end{figure*}

Here, we compare our spectral function with the results obtained for model 2c in Ref.\,\cite{Brandt:2015aqk}. 
The other models of that work have different features in certain regions of $\omega$, but the same overall behavior. 
In Fig.\,\ref{fig:full.spec.comp} we show 
the full (nonsubtracted) spectral functions for four different temperatures. 
It is seen in these figures that while the details of model 2c differ from our spectral function, the 
general structure is the same. The biggest difference between our spectral function and those of Ref.\,\cite{Brandt:2015aqk} 
can be found for the $T = 0.8\,T_c$ case in the region around the $\rho$ meson peak, where all models of Ref.\,\cite{Brandt:2015aqk} 
are close to 0, while our spectral function is smoothly connected to the continuum. The true spectral function in this region likely  
lies between these two extremes, as the $\rho$ meson in reality has quite a large width and is placed on top of a smooth $\pi \pi$ continuum, 
but still is the dominating structure below energies of about $1$ GeV (see, for instance, Fig.\,1 of Ref.\,\cite{Kwon:2008vq}). 
%To compare how well the above spectral functions reproduce the lattice data of Ref.\,\cite{Brandt:2015aqk}, 
%they are both shown as red and blue data points in Fig.\,\ref{fig:comp.latt}, where it is observed that the spectral function of 
%Ref.\,\cite{Brandt:2015aqk} reproduces the lattice data of the vacuum ensemble slightly better than ours for large $\tau$ 
%values. For the finite temperature ensembles, the lattice data reproduction of our spectral function is comparable to Model 2c 
%(as well as the other models) of Ref.\,\cite{Brandt:2015aqk}. 

%%%%%%%%%%%%%%%%%%%%%%%%%%%%%%%%%%%%%%%%%%%
\subsection{Evaluation of physical quantities}
\begin{figure}
\begin{center}
%\vspace{-1.0cm}
%\hspace*{1.2cm}
\includegraphics[width=0.49\textwidth]{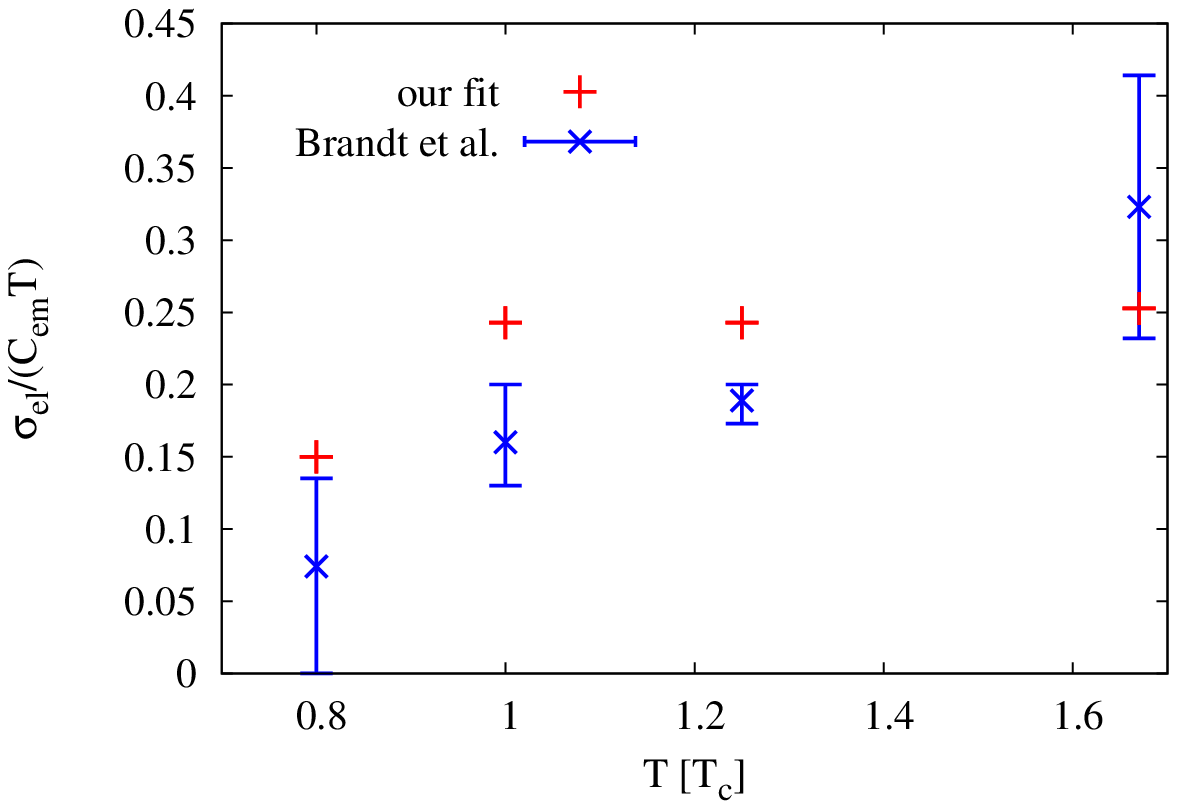}
%\hspace*{1.2cm}
\includegraphics[width=0.49\textwidth]{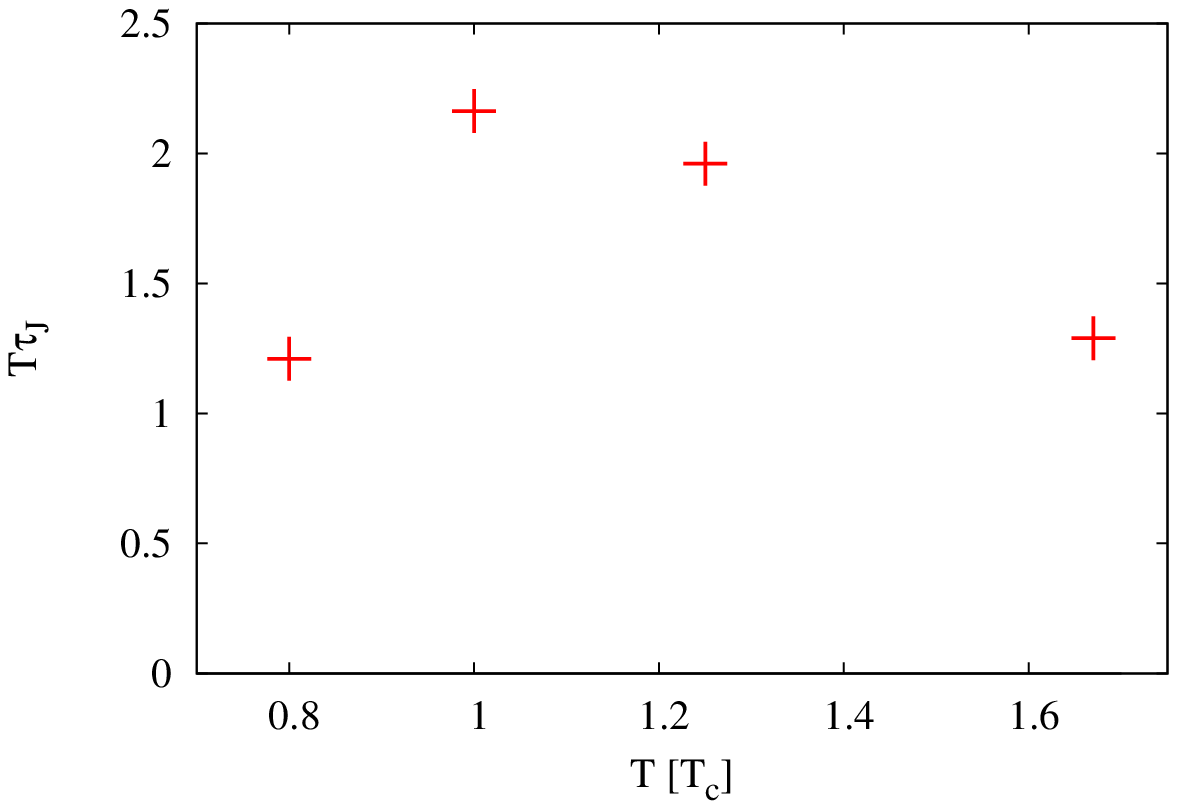}
%\vspace{0.5cm}
\caption{The electric conductivity $\sigma_{\mathrm{el}}$ and the second order transport coefficient $\tau_{J}$ as functions of 
the temperature $T$. Both quantities are made dimensionless by multiplying appropriate powers of $T$.
For $\sigma_{\mathrm{el}}$, the values obtained in Ref.\,\cite{Brandt:2015aqk} are also shown. Here, we have 
for simplicity combined the various errors quoted in Ref.\,\cite{Brandt:2015aqk} into a single one for each data point.}  
\label{fig:low.en.const}
\end{center} 
\end{figure}

From the above fit results, we can determine the electric conductivity from the numerical values of $A_{T}/\Gamma_{T}$. 
In our parametrization, it is given as 
\begin{align}
\frac{\sigma_{\mathrm{el}}}{C_{\mathrm{em}}} = \frac{1}{C_{\mathrm{em}}} \lim_{\omega \to 0} \frac{\rho(\omega,T)}{\omega} = \frac{1}{3} \frac{A_T}{\Gamma_T}.  
\label{eq:elec.conduct}
\end{align} 
From the values of Table\,\ref{tab:fit.values}, $\sigma_{\mathrm{el}}$ is obtained as shown in the second column of Table\,\ref{tab:le.constants} 
and on the upper panel of Fig.\,\ref{fig:low.en.const}. 
We however emphasize once again here, that other fits with similar $\chi^2$ values but rather different electric conductivities 
are possible and that the numbers shown here just represent one of many possible solutions. 
On the upper panel of Fig.\,\ref{fig:low.en.const}, we furthermore show the results obtained in Ref.\,\cite{Brandt:2015aqk} for 
comparison. It is seen that with the exception of the point at $T = 1.67\,T_c$, the results are not consistent, even though 
they show the same general tendency. This once more indicates that the systematic uncertainty in the evaluation of this quantity 
is still rather large. 

Next, we can now use sum rule 3 to estimate the value of the second order transport coefficient $\tau_{J}$, as all other ingredients in that 
sum rule are known. The results of such a computation are found in the third column of Table\,\ref{tab:le.constants} 
and on the lower plot of Fig.\,\ref{fig:low.en.const}. 
\begin{table}
\renewcommand{\arraystretch}{1.3}
\begin{center}
\caption{The electric conductivity $\sigma_{\mathrm{el}}$ and the second order transport coefficient $\tau_{J}$ at various 
temperatures, as obtained from the fit to lattice QCD data.} 
\label{tab:le.constants}
\begin{tabular}{ccc}  
\hline 
$T$ & $\sigma_{\mathrm{el}}/(C_{\mathrm{em}}T)$ & $T\tau_{J}$  \\ \hline
$0.8\,T_c$ & 0.150 & 1.21 \\
$1.0\,T_c$ & 0.243 & 2.16 \\
$1.25\,T_c$ & 0.243 &  1.96 \\
$1.67\,T_c$ & 0.253 &  1.29 \\
\hline
\end{tabular}
\end{center}
\end{table}
 It is seen in this figure that $\tau_{J}$ exhibits quite an interesting behavior as 
a function of $T$. Namely, it increases for temperatures below $T_c$, takes a maximum at $T = T_c$ and 
then decreases again for temperatures above $T_c$. It remains to be seen whether 
this behavior is an artifact of our fit and/or our parametrization of the spectral function or if it is a real physical 
effect. 

\begin{figure}
\begin{center}
%\vspace{-1.0cm}
%\hspace*{1.2cm}
\includegraphics[width=0.49\textwidth]{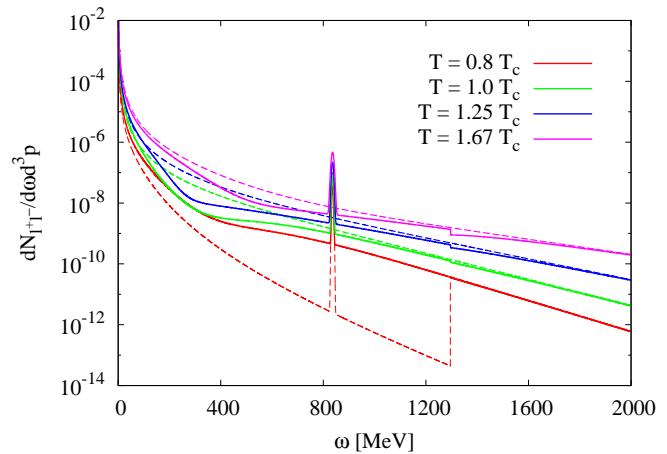}
%\vspace{0.5cm}
\caption{The thermal dilepton rate for two-flavor QCD, extracted from our fit to lattice QCD data of Ref.\,\cite{Brandt:2015aqk}. 
The thick solid lines show the result of our fit, while the thin dashed lines correspond to model 2c of Ref.\,\cite{Brandt:2015aqk}.}  
\label{fig:dilepton.rate}
\end{center} 
\end{figure}

As a last result, we give the thermal dilepton rate $dN_{l^{+}l^{-}}/d\omega d^3 \vp$ for vanishing 
spatial momentum ($|\vp| = 0$), which can be easily obtained from the relation between the spectral function and 
the dilepton rate: 
\begin{align}
\frac{dN_{l^{+}l^{-}}}{d\omega d^3 \vp}(|\vp|=0)  = \frac{\alpha_{\mathrm{em}}^2}{\pi^3 \omega^2} 
\frac{\rho(\omega,T)}{e^{\omega/T} - 1}. 
\end{align}
The results are shown in Fig.\,\ref{fig:dilepton.rate}, where we have adjusted the horizontal axis to physical units (MeV) and 
where we also show the corresponding model 2c results of Ref.\,\cite{Brandt:2015aqk}. 
We see that our result is larger than the one in Ref.\,\cite{Brandt:2015aqk}, especially at $T=0.8T_c$.
This difference can be understood from the absence of the spectrum near the vector meson peak in Ref.\,\cite{Brandt:2015aqk}, 
which was discussed in the previous subsection. 

%%%%%%%%%%%%%%%%%%%%%%%%%%%%%%%%%%%%%%%%%%%
\section{Summary and Concluding Remarks}
\label{sec:summary}
In the first part of this paper, we derived and discussed five exact sum rules (two for the transverse and three for the longitudinal parts) 
for the vacuum-subtracted spectral functions of the vector channel at finite temperature, which are 
determined completely by the UV and IR behavior of the vector correlator. 
The UV part can be obtained from the OPE, while the 
IR behavior is accurately described by hydrodynamics. 
The sum rules are valid for nonzero momentum $\vp$, which should however be 
small enough such that the hydrodynamic description of the vector correlator in the IR regime can be trusted. 
In the limit $|\vp| \to 0$, transverse and longitudinal parts approach each other such that the sum rules which 
we have already derived in Ref.\,\cite{Gubler:2016hnf} remain. 

In perturbation theory, it has been known that three (four) distinct structures emerge in the transverse (longitudinal) channel: 
a transport peak, an exponentially suppressed continuum, and a power suppressed UV tail in both channels, and a diffusion peak in the longitudinal channel. 
All these structures can, in principle, contribute to the various sum 
rules and therefore need to be taken into account to test their validity. 
Doing this, we found that the sum rules are indeed satisfied in the weak coupling regime. This exercise 
also gives a rough idea about how the different parts of the full spectral function can be expected to 
contribute in different ways to each sum rule. 

In the second part of the paper, we have employed recent two-flavor dynamical lattice QCD data at almost zero and 
finite temperature and at zero momentum to perform a spectral fit, in which two sum rules (1 and 2) are used as constraints to reduce the 
number of parameters to be fitted. The third sum rule (3) in turn enables us to extract the value of the 
second order transport coefficient $\tau_J$. 
We note that the second sum rule was not used in our previous work~\cite{Gubler:2016hnf}.
It, however, needs to be emphasized here that we found the fit not to be completely stable in the 
sense that we confirmed the existence of several local minima with comparable $\chi^2$ values, which 
means that more data points with increased precision will be needed for uniquely determining the true 
shape of the spectral function. 
Nevertheless, we succeeded in demonstrating that the sum rules can be used to improve the lattice QCD data analysis.

In our fit, we employed lattice data at zero momentum and 
have therefore used the three sum rules already derived in Ref.\,\cite{Gubler:2016hnf}. 
Once lattice data at nonzero momentum are available, it would be interesting to apply the sum rules 
derived in this paper to their analysis. 
Also, once the lattice QCD analysis at the physical point 
and in the continuum limit becomes 
available, it will be possible to use the phenomenological form 
of the vector spectral function at $T=0$ obtained from experiment.
We leave these topics for future work. 

%%%%%%%%%%%%%%%%%%%%%%%%%%%%%%%%%%%%%%%%%%%%%%%%%%%
\section*{Acknowledgements}

The research of D. S. is supported by the Alexander von Humboldt Foundation. 
The research of P.G. is supported 
by the Mext-Supported Program for the Strategic 
Foundation at Private Universities, ``Topological Science" (No. S1511006). 
We thank Bastian B. Brandt for fruitful discussions.

%%%%%%%%%%%%%%%%%%%%%%%%%%%%%%%%%%%%%%%%%%%%
\appendix
\section{Evaluation of the transport peak at weak coupling}
\label{app:transport-peak}

In this appendix, we evaluate the transport peak appearing in the spectral function by using the Boltzmann equation 
in the relaxation time approximation for massless quarks. 
This appendix is in essence a recapitulation of the literature, for example Refs.~\cite{Satow:2014lva, Hong:2010at}. 
The Boltzmann equation reads 
\begin{align}
\begin{split} 
&v\cdot\partial_X n_{\pm f}(\vk, X) 
\pm eq_f\left(\vE+\vv\times\vB\right)(X) \cdot\nabla_{\vk} n_{\pm f}(\vk, X)\\
&= C[n_{\pm f}], 
\end{split}
\end{align}
where $n_{\pm f}(\vk, X)$ is the distribution function for the quark (antiquark) with momentum $\vk$ at point $X$, and $v^\mu\equiv (1,\vv)$ with $\vv\equiv \vk/|\vk|$. 
$C[n_{\pm f}]$ represents the collision effect among the quarks, which is given later. 

Now we consider the situation in which the system at equilibrium is disturbed by weak external EM fields, 
so that the distribution function slightly deviates from the equilibrium, $n_{\pm f}(\vk, X)=\nf(|\vk|)+\delta n_{\pm f}(\vk, X)$. 
By linearizing the Boltzmann equation in terms of $\delta n_{\pm f}(\vk, X)$ and EM fields, we get
\begin{align}
\label{eq:EOM-1}
\begin{split} 
v\cdot\partial_X \delta n_{\pm f}(\vk, X) 
\pm eq_f \vE(X) \cdot\vv \nf'(|\vk|)
&= \delta C[n_{\pm f}], 
\end{split}
\end{align}
where the magnetic field term disappears due to the isotropy of the system at equilibrium.
$\delta C[n_{\pm f}]$ is a linearized form of $C[n_{\pm f}]$, whose expression reads
\begin{align}
\delta C[n_{\pm f}]
&= -\tau^{-1} \left(\delta n_{\pm f}(\vk, X) 
\mp\nf'(|\vk|) \delta\mu_f(X)
\right), 
\end{align}
in the relaxation time approximation.
Here we have introduced the relaxation time $\tau\sim(g^4T)^{-1}$, and the shift of the chemical potential ($\delta\mu_f$) caused by the EM fields. 
The second term in the expression above is necessary, since the deviation of the distribution created by the shift of the chemical potential does not relax. 
The shift of the chemical potential is determined by the conservation law of particle number, 
\begin{align}
0&=\int\frac{d^3\vk}{(2\pi)^3}\sum_{s=\pm 1} s  \delta C[n_{s f}] ,
\end{align}
which reduces to
\begin{align}
\label{eq:EOM-2}
\delta\mu_f(X)
&= -\frac{1}{\chi}
\int\frac{d^3\vk}{(2\pi)^3}\sum_{s=\pm 1} s  \delta n_{s f}(\vk, X).
\end{align}
Here, $\chi\equiv T^2/6$. 
From Eqs.~(\ref{eq:EOM-1}) and (\ref{eq:EOM-2}), we obtain the solutions, 
\begin{align}
\delta\mu_f(p)
&= ieq_f\frac{\vE(p)\cdot \hat{\vp}}{|\vp|} \frac{1-(\omega+i\tau^{-1})A(p)}{1-i\tau^{-1}A(p)}, \\
\label{eq:Boltzmanneq-solution}
\delta n_{s f}(\vk, p)
&= -is\nf'(|\vk|)\frac{eq_f\vE(p)\cdot\vv-\tau^{-1}\delta\mu_f(p)}{v\cdot p+i\tau^{-1}},
\end{align}
where we have performed the Fourier transformation $(X\rightarrow p)$ and introduced $A(p)\equiv \ln[(\omega+|\vp|+i\tau^{-1})/(\omega-|\vp|+i\tau^{-1})]/(2|\vp|)$.

The induced current is given by 
\begin{align}
j^\mu(p)
&= 2e\sum_f q_f\Nc\int\frac{d^3\vk}{(2\pi)^3} v^\mu
\sum_{s=\pm 1} s\delta n_{sf}(\vk,p).
\end{align}
By using this expression in momentum space and the linear response theory relation of  Eq.\,(\ref{eq:linear-response}), 
Eq.~(\ref{eq:Boltzmanneq-solution}) leads to the retarded Green function $G^R_{\mu\nu}(p)$, as is shown below.
We note that $j^0=-2e\sum_f q_f\Nc\chi\delta\mu_f$, which indicates that $\chi$ is essentially the susceptibility.

%%%%%%%%%%%%%%%%%%%%
\subsection{Transverse channel}

The transverse component of $G^R$ is given by 
\begin{align}
\begin{split}
G_T(p)
&=4 \Cem\Nc\frac{1}{(2\pi)^3}
\int^{2\pi}_0 d\phi \int^1_{-1} d\cos\theta \int^\infty_0 d|\vk| |\vk|^2 \\
&~~~\times\nf'(|\vk|)\frac{\omega\sin^2\theta\cos^2\phi}{\omega-|\vp|\cos\theta+i\tau^{-1}} \\
&= -\Cem\Nc\chi\frac{\omega}{\vp^2} \\
&~~~\times \left[\omega+i\tau^{-1}-A(p)(p^2+2i\tau^{-1}\omega+(i\tau^{-1})^2)\right].
\end{split}
\label{eq:GTfull}
\end{align} %\chi is used as exact suscptibility in hydro. So, we should use other symbol?

In the hydro limit, $\omega,|\vp|\ll\tau^{-1}$, this reduces to 
\begin{align}
\begin{split}
G_T(p)
&\simeq \Cem\Nc\chi \frac{2}{3}\tau \omega
 \left[i-\tau\omega\right],
\end{split} 
\end{align}
where we have used 
\begin{align}
\begin{split}
A(p) & \simeq \frac{1}{i\tau^{-1}}
\left[1-\frac{\omega}{i\tau^{-1}}+\frac{3\omega^2+\vp^2}{3(i\tau^{-1})^2}
-\frac{\omega(\omega^2+\vp^2)}{(i\tau^{-1})^3}
\right].
\end{split}
\end{align}
By comparing this expression with the hydro result of Eq.\,(\ref{eq:hydro-GR-T}), we derive the following expression for the transport coefficients:
\begin{align}
\sigma &= \Cem\Nc\chi \frac{2}{3}\tau
= \Cem\Nc\frac{T^2}{9}\tau,\\
\tau_J&=\tau, \\
\kappa_B&=0.
\end{align}

Next, we obtain the spectral function for small $|\vp|$. 
Expanding Eq.\,(\ref{eq:GTfull}) in terms of $|\vp|$, we derive 
\begin{align}
\begin{split}
G_T(p)
&\simeq -\Cem\Nc\chi \frac{2}{3}\frac{\omega}{\omega+i\tau^{-1}} 
 \left[1+\frac{1}{5}\frac{\vp^2}{(\omega+i\tau^{-1})^2}\right],
\end{split} 
\label{eq:GTexpanded}
\end{align}
where we have used 
\begin{align}
\begin{split}
& A(p) \simeq \\
& \frac{1}{\omega+i\tau^{-1}}
\left[1+\frac{1}{3}\left(\frac{|\vp|}{\omega+i\tau^{-1}}\right)^2
 +\frac{1}{5}\left(\frac{|\vp|}{\omega+i\tau^{-1}}\right)^4\right].
\end{split}
\end{align}
The imaginary part of Eq.\,(\ref{eq:GTexpanded}) reads
\begin{align}
\label{eq:transport-T}
\begin{split}
\rho_T(p)
&\simeq \Cem\Nc\chi \frac{2}{3}  \frac{\tau^{-1} \omega}{\omega^2+\tau^{-2}}
 \left[1 +\frac{\vp^2}{5} \frac{(3\omega^2-\tau^{-2})}{(\omega^2+\tau^{-2})^2}\right].
\end{split} 
\end{align}

%%%%%%%%%%%%%%%%%%%%
\subsection{Longitudinal channel}

The longitudinal component of $G^R$ is given by
\begin{align}
\begin{split}
G^R_{00}(p)
&= 2\Cem\Nc\chi\frac{1-(\omega+i\tau^{-1})A(p)}{1-i\tau^{-1}A(p)}.
\end{split}
\label{eq:G00full}
\end{align}

In the hydro limit, $\omega,|\vp|\ll\tau^{-1}$, this reduces to 
\begin{align}
\label{eq:G00-hydrolim}
\begin{split}
G^R_{00}(p)
%&\simeq 
%i\vp^2 \frac{2\Cem\Nc\chi \tau}{3}
%\frac{1+2i\tau\omega}
%{(\omega+i\tau\vp^2/3+i\tau \omega^2-\tau^2\omega\vp^2)} \\
&\simeq i\vp^2 \frac{2\Cem\Nc\chi \tau}{3}
\frac{1}
{(\omega+i\tau\vp^2/3)},
\end{split} 
\end{align}
where we have retained only the leading order terms.
Comparing this expression with the result of hydrodynamics (\ref{eq:hydro-GR-L}), we get
\begin{align}
D&= \frac{\tau}{3}, 
\end{align}
and also confirm the Einstein relation, 
\begin{align}
\sigma
&= 2\Cem\Nc \chi D,
\end{align}
where the factor 2$\Nc$ originates from the spin and color degrees of freedom of the quark.

Now, we can obtain the spectral function for small $|\vp|$. 
Expanding Eq.\,(\ref{eq:G00full}) in terms of $|\vp|$, we derive 
\begin{align}
\begin{split}
G^R_{00}(p)
&\simeq -\frac{2}{3}\Cem\Nc\chi \frac{|\vp|^2}{\omega+i\tau^{-1}}
\frac{1}{\omega} \\
&~~~\times \left[1+\left(\frac{3}{5}+\frac{i\tau^{-1}}{3\omega}\right)
\left(\frac{|\vp|}{\omega+i\tau^{-1}}\right)^2
\right].
\end{split} 
\end{align}
The imaginary part of the above expression reads
\begin{align}
\label{eq:transport-00}
\begin{split}
\rho^{00}(p)
&\simeq 
\frac{2}{3}\Cem\Nc\chi \frac{\vp^2}{\omega} \frac{\tau^{-1}}{\omega^2+\tau^{-2}} \\
&~~~\times \left[1
+\vp^2 \frac{2}{5}\left(\tau^{-2}+\frac{11}{3}\omega^2\right)
\frac{1}{(\omega^2+\tau^{-2})^2}
\right].
\end{split} 
\end{align}

%%%%%%%%%%%%%%%%%%%%%%%%%%%%%%%%%%%%%%%%%%%%
\section{Evaluation of the continuum at weak coupling}
\label{app:continuum}

In this appendix, we evaluate the continuum 
in the weak coupling and massless limit. 
In the free limit, the Green function of the EM current can be calculated by using Wick's theorem as~\cite{Altherr:1989jc}
\begin{align}
\begin{split} 
G^R_{\mu\nu}(x)
&= i\theta(t) C_{\text{em}}\Nc
\Tr\Bigl[\gamma_\mu S^>(x)\gamma_\nu S^<(-x) \\
&~~~-\gamma_\mu S^<(x)\gamma_\nu S^>(-x)
\Bigr],
\end{split}
\end{align}
where $S^>(x)\equiv \langle \psi(x)\overline{\psi}(0) \rangle$ and $S^<(x)\equiv \langle \overline{\psi}(0) \psi(x)\rangle$.
Here, we have omitted the flavor indices for simplicity. 
By performing the Fourier transformation and taking the imaginary part, the spectral function reads
\begin{align}
\begin{split} 
\rho_{\mu\nu}(p)
&= \frac{1}{2} C_{\text{em}}\Nc \int \frac{d^4k}{(2\pi)^4}
\Tr\Bigl[\gamma_\mu S^>(k)\gamma_\nu S^<(k-p) \\
&~~~-\gamma_\mu S^<(k)\gamma_\nu S^>(k-p)
\Bigr]\\
&=2 C_{\text{em}}\Nc \int \frac{d^4k}{(2\pi)^4}
\Tr\Bigl[\gamma_\mu \Slash{k}\gamma_\nu (\Slash{k}-\Slash{p}) \Bigr] \\
&~~~\times \rho^0(k) \rho^0(k-p)
[\nf(k^0-\omega)-\nf(k^0)],
\end{split}
\label{eq:rhomunu1}
\end{align}
where we have used $S^>(k)= \Slash{k}2\rho^0(k)[1-\nf(k^0)]$ and $S^<(k)=\Slash{k}2\rho^0(k)\nf(k^0)$, 
and introduced the free quark spectral function as $\rho^0(k)\equiv \pi\sgn(k^0)\delta(k^2)$. 
The two delta functions can be written as 
\begin{align}
\label{eq:deltafunc-1}
\delta(k^2)&=\sum_{s=\pm 1}\frac{\delta(k^0-s|\vk|)}{2|\vk|},\\
\nonumber
\delta([k-p]^2)&= \frac{1}{2|\vk||\vp|}\delta\left[\cos\theta-\frac{2k^0\omega-p^2}{2|\vk||\vp|}\right] \\
\label{eq:deltafunc-2}
&~~~\times\theta\left[-p^2\left\{(k^0)^2-k^0\omega+\frac{p^2}{4}\right\}\right],
\end{align}
where we adopted the standard polar coordinate, by assigning $\vp$ 
to point into the $z$-direction. 
We also note that $\theta$ in the first line is the angle between $\vk$ and $\vp$ while that in the second line is a step function.
Therefore, the expression of Eq.\,(\ref{eq:rhomunu1}) becomes  
\begin{align}
\begin{split}
\rho_{\mu\nu}(p)
&= C_{\text{em}}\Nc 
\frac{1}{8\pi^2|\vp|}
\int^{2\pi}_0 d\phi \int^\infty_0 d|\vk| \sum_{s=\pm 1} \\
&~~~\times \Bigl[k_\mu (k-p)_\nu+k_\nu (k-p)_\mu+g_{\mu\nu} k\cdot p \Bigr] \\
&~~~\times  \sgn(s|\vk|-\omega)
s \theta\left[-\left\{(k^0)^2-k^0\omega+\frac{p^2}{4}\right\}\right]\\
&~~~\times [\nf(s|\vk|-\omega)-\nf(s|\vk|)],
\end{split}
\label{eq:rhomunu2}
\end{align}
where the values for $k^0$ and $\cos\theta$ are determined by the delta functions of Eqs.~(\ref{eq:deltafunc-1}) and (\ref{eq:deltafunc-2}), 
and we have used $p^2>0$, which is justified because we consider the case that $\omega \sim T\gg |\vp|$. 
From now on, we focus on the case for which $\omega>0$. 
Then, only the contribution with $s=+1$ remains, and the step function restricts the range of $|\vk|$ to $k_-<|\vk|<k_+$, where $k_\pm\equiv (\omega\pm|\vp|)/2$.
Thus, Eq.\,(\ref{eq:rhomunu2}) becomes
\begin{align}
\label{eq:rho-continuum}
\begin{split}
\rho_{\mu\nu}(p)
&= -C_{\text{em}}\Nc 
\frac{1}{8\pi^2|\vp|}
\int^{2\pi}_0 d\phi \int^{k_+}_{k_-} d|\vk|  \\
&~~~\times \Bigl[k_\mu (k-p)_\nu+k_\nu (k-p)_\mu+g_{\mu\nu} k\cdot p \Bigr] \\
&~~~\times   [1-\nf(-|\vk|+\omega)-\nf(|\vk|)],
\end{split}
\end{align}
where we have used $\omega-|\vk|>(\omega-|\vp|)/2>0$.
From the distribution factor $[1-\nf(-|\vk|+\omega)-\nf(|\vk|)]=[1-\nf(-|\vk|+\omega)][1-\nf(|\vk|)]-\nf(-|\vk|+\omega)\nf(|\vk|)$, 
we see that the physical processes corresponding to this expression are the quark antiquark 
pair-creation process and its inverse. 
%should we draw the cutted diagram?

%%%%%%%%%%%%%%%%%%%%
\subsection{Transverse channel}

$\delta\rho^T(p)$ can be evaluated by setting $\mu=\nu=1$ and subtracting the $T=0$ part from Eq.~(\ref{eq:rho-continuum}):
\begin{align}
\begin{split}
\delta\rho_{T}(p)
&= C_{\text{em}}\Nc 
\frac{1}{8\pi^2|\vp|}
\int^{2\pi}_0 d\phi \int^{k_+}_{k_-} d|\vk|  \\
&~~~\times \Bigl[2\vk^2\cos^2\phi(1-\cos^2\theta)-|\vk|\omega+|\vk||\vp|\cos\theta \Bigr] \\
&~~~\times   [\nf(-|\vk|+\omega)+\nf(|\vk|)]\\
&= -C_{\text{em}}\Nc 
\frac{p^2}{4\pi|\vp|^3}
\int^{|\vp|/2}_{-|\vp|/2} d|\vk|  
\left(\frac{\vp^2}{4}+\vk^2\right) \\
&~~~\times \left[\nf\left(\frac{\omega}{2}-|\vk|\right)+\nf\left(\frac{\omega}{2}+|\vk|\right)\right],
\end{split}
\end{align}
where in the last line, we have used $\cos\theta=(2|\vk|\omega-p^2)(2|\vk||\vp|)$ and have changed the integration variable as $|\vk|\rightarrow |\vk|-\omega/2$.

We can safely expand this in terms of $|\vk|/\omega$, because $|\vk|\simeq |\vp|\ll \omega$, which leads to 
\begin{align}
\label{eq:continuum-T}
\begin{split}
& \delta\rho_{T}(p) \\ 
&\simeq - C_{\text{em}}\Nc 
\frac{\omega^2}{\pi|\vp|^3}\left(1-\frac{\vp^2}{\omega^2}\right)
\int^{|\vp|/2}_{0} d|\vk|  
\left(\frac{\vp^2}{4}+\vk^2\right) \\
&~~~\times \left[\nf\left(\frac{\omega}{2}\right)+\frac{\vk^2}{2}\nf''\left(\frac{\omega}{2}\right)\right]\\
&\simeq - C_{\text{em}}\Nc 
\frac{\omega^2}{6\pi}\left(1-\frac{\vp^2}{\omega^2}\right) 
 \left[\nf\left(\frac{\omega}{2}\right)+\frac{\vp^2}{20}\nf''\left(\frac{\omega}{2}\right)\right]\\
 &\simeq -C_{\text{em}}\Nc 
\frac{\omega^2}{6\pi}
 \Biggl[\nf\left(\frac{\omega}{2}\right) \\
&~~~ +\vp^2\left\{\frac{1}{20}\nf''\left(\frac{\omega}{2}\right)
 -\frac{1}{\omega^2}\nf\left(\frac{\omega}{2}\right)\right\}\Biggr] .
\end{split}
\end{align}

%%%%%%%%%%%%%%%%%%%%
\subsection{Longitudinal channel}

By setting $\mu=\nu=0$ in Eq.~(\ref{eq:rho-continuum}), we get 
\begin{align}
\begin{split}
\delta\rho_{00}(p)
&= C_{\text{em}}\Nc 
\frac{1}{2\pi|\vp|}
\int^{|\vp|/2}_{-|\vp|/2} d|\vk|  
\left[ \vk^2- \frac{\vp^2}{4} \right] \\
&~~~\times   \left[\nf\left(\frac{\omega}{2}-|\vk|\right)+\nf\left(\frac{\omega}{2}+|\vk|\right)\right],
\end{split}
\end{align}
where we have changed the integration variable as before.

Expanding the integrand in terms of $|\vp|/\omega$, we derive 
\begin{align}
\label{eq:continuum-00}
\begin{split}
\delta\rho_{00}(p)
&= C_{\text{em}}\Nc 
\frac{2}{\pi|\vp|}
\int^{|\vp|/2}_{0} d|\vk|  
\left[ \vk^2- \frac{\vp^2}{4} \right] \\
&~~~\times \left[\nf\left(\frac{\omega}{2}\right)+\frac{\vk^2}{2}\nf''\left(\frac{\omega}{2}\right)\right]\\
&= -C_{\text{em}}\Nc  \frac{\vp^2}{6\pi} 
 \left[\nf\left(\frac{\omega}{2}\right)+\frac{\vp^2}{40}\nf''\left(\frac{\omega}{2}\right)\right] .
\end{split}  
\end{align}

%%%%%%%%%%%%%%%%%%%%%%%%%%%%%%%%%%%%%%%%%%%%
\section{Evaluation of the UV tail at weak coupling}
\label{app:UVtail}

In this appendix, we briefly recapitulate the derivation of the UV tail in the EM current spectral function from the OPE~\cite{CaronHuot:2009ns}.
The UV behavior of the EM current retarded correlator is described by the OPE of Eqs.\,(\ref{eq:OPE-wo-RG-T}) and (\ref{eq:OPE-wo-RG-L}).
Among the three terms in these expressions, only $\langle T^{00}_f\rangle$ is not RG invariant.
This operator yields imaginary parts of the retarded correlator, as can be understood as follows: 
The scaling relation (\ref{eq:scaling-T'-Ttilde}) can be rewritten as
\begin{align}
\begin{split} 
T'{}^{00}_{f}(\kappa)
&\simeq  T'{}^{00}_{f}(\kappa_0)
+a'\ln\left(\frac{\kappa^2_0}{\kappa^2}\right)
\frac{b_0}{4\pi}\alpha_sT'{}^{00}_{f} ,\\
 \tilde{T}^{00}(\kappa)
&\simeq  \tilde{T}^{00}(\kappa_0)
+\tilde{a}\ln\left(\frac{\kappa^2_0}{\kappa^2}\right)
\frac{b_0}{4\pi}\alpha_s\tilde{T}^{00}_{f},
\end{split} 
\end{align}
when $\kappa$ is close to $\kappa_0$.
It was shown in Ref.\,\cite{CaronHuot:2009ns} 
that the factor $\ln(\kappa^2_0/\kappa^2)$ generates an imaginary contribution $i\pi$, due to the analytic continuation to the real time. 
Following this prescription, the imaginary parts of the retarded correlators (\ref{eq:OPE-wo-RG-T}) and (\ref{eq:OPE-wo-RG-L}) read
\begin{align}
\nonumber
\delta \rho_T(p) 
&=e^2\sum q^2_f \frac{8}{9}\frac{\omega^2+\vp^2}{(p^2)^2}
\alpha_s(\omega) \\
\label{eq:deltarho_T_UVtail}
&~~~\times  \left( 2\Cf \delta\left\langle T'{}^{00}_{f}(\omega)  \right\rangle
+\frac{1}{\Nf}  \delta\left\langle \tilde{T}^{00}(\omega) \right\rangle \right) ,\\
\nonumber
\delta \rho_{00}(p)
&=  e^2\sum q^2_f \frac{8}{9}\frac{\vp^2}{(p^2)^2}
\alpha_s(\omega) \\
&~~~\times  \left( 2\Cf \delta\left\langle T'{}^{00}_{f}(\omega)  \right\rangle
+\frac{1}{\Nf}  \delta\left\langle \tilde{T}^{00}(\omega) \right\rangle \right) .
\end{align}
We note that this expression is valid when the OPE is reliable ($\omega\gg T,\Lambda_{\text {QCD}}$).

In the chiral and weak coupling limits, the operator expectation values at the renormalization scale $\kappa_0\sim T$ read
\begin{align}
\label{eq:T00f-free}
\langle T^{00}_f\rangle&= \Nc\frac{7\pi^2T^4}{60},\\
\label{eq:T00g-free}
\langle T^{00}_g\rangle&=2\Cf \Nc \frac{\pi^2T^4}{15},
\end{align}
which, by using the scaling relation of Eq.\,(\ref{eq:scaling-T'-Ttilde}), leads to 
\begin{align}
\nonumber
\delta \rho_T(p) 
&=\Cem \frac{1}{\omega^2} 
\left(1+3\frac{\vp^2}{\omega^2}\right)
\alpha_s(\kappa_0) \Nc\Cf\frac{4\pi^2T^4}{27} \\
\label{eq:UVtail-T}
&~~~\times  \left[\frac{\ln\left(\kappa_0/\Lambda_{\text{QCD}} \right)}{\ln\left(\omega/\Lambda_{\text{QCD}} \right)}\right]^{\tilde{a}+1}  ,\\
\nonumber
\delta \rho_{00}(p)
&=\Cem \frac{\vp^2}{\omega^4} 
\left(1+2\frac{\vp^2}{\omega^2}\right)
\alpha_s(\kappa_0) \Nc\Cf \frac{4\pi^2T^4}{27}\\
\label{eq:UVtail-00}
&~~~\times  \left[\frac{\ln\left(\kappa_0/\Lambda_{\text{QCD}} \right)}{\ln\left(\omega/\Lambda_{\text{QCD}} \right)}\right]^{\tilde{a}+1}. 
\end{align} %we can not eliminate kappa_0 dependence? should we choose as \kappa_0=e\LambdaQCD?
Here, we have retained terms up to next-to-leading order in the small $|\vp|$ expansion.

%%%%%%%%%%%%%%%%%%%%%%%%%%%%%%%%%%%%%%%%%%%%%%%%

\end{document}